\PassOptionsToPackage{table}{xcolor}
\documentclass[twocolumn]{aastex631}

\usepackage[T1]{fontenc}
\usepackage{amsmath}
\usepackage{multirow}
\usepackage{booktabs}
\usepackage{tabularx}
\usepackage{graphicx}
\usepackage{xcolor}

\begin{document}

\title{The Effects of Complex Accretion Disk Geometry on Broadened Iron K$\alpha$ Lines}

\author[0000-0002-8108-6904]{William Surgent}
\affiliation{Department of Physics, University of California, Santa Barbara, CA 93106, USA}
\affiliation{Department of Physics, Stanford University, Stanford, California 94305, USA}
\affiliation{Kavli Institute for Particle Astrophysics \& Cosmology (KIPAC), Stanford University, Stanford, CA 94305, USA}

\author[0000-0002-4794-5998]{Daniel R. Wilkins}
\affiliation{The Ohio State University, Department of Astronomy, 4055 McPherson Laboratory, 140 W 18th Ave, Columbus, OH 43210, USA}
\affiliation{Kavli Institute for Particle Astrophysics \& Cosmology (KIPAC), Stanford University, Stanford, CA 94305, USA}



\begin{abstract}
X-rays are emitted from the corona above the orbiting matter of the accretion disk and travel either directly to us or illuminate the disk. This illumination of the inner disk is enhanced by gravitational light bending, which focuses the rays towards the black hole and therefore towards the inner radii of the disk. These rays that hit the inner radii are reflected back to us, and we observe them in the X-ray reflection spectrum. In this work, we create novel general-relativistic ray-tracing simulations to investigate the effects of altering the geometry of the accretion disks of black holes on the most dominant part of the reflection spectrum, the iron K$\alpha$ line. Work demonstrating the effect of disk geometry on the iron line has been performed, though many previous analyses have assumed a simplistic system, consisting of a point-source corona with a flat and infinitesimally thin accretion disk. We extend these models to more realistic accretion disk approximations. These include a constant-aspect-ratio disk, a radiation-pressure-dominated Shakura-Sunyaev disk, an expanded inner disk that has a nonnegligible scale height in its inner regions due to radiation pressure, as well as various warped-disk configurations. Using measurement uncertainties from XRISM, we find that nonnegligible thickness in accretion disks underestimates the black hole spin, coronal height, and inclination angle if fitted with a flat-disk model. The warped-disk model could not be fit with the flat-disk approximation. 
\end{abstract}



\section{Introduction} \label{sec:intro}

Black holes are interesting phenomena predicted by Einstein's theory of general relativity, whose interiors are causally isolated from the rest of the Universe. Matter falling into black holes can fuel some of the most powerful objects in the Universe: active galactic nuclei (AGNs). Around the black hole is an accretion disk of matter, above which there is a luminous corona consisting of hot and energized particles that emits X-rays. The corona is believed to be heated by the reconnection of magnetic fields, which accelerate particles in the region. One possible mechanism is the differential rotation of the material across various radii in the accretion disk, in addition to convection motions within it, which stretches and enhances its magnetic field lines. The resulting loops of magnetic field create a corona around the accretion disk with a very high temperature of over $10^8$ K \citet{1979ApJ}. The corona emits X-rays. A portion of these X-rays escape from the black hole and we observe them as the (power law) continuum. Another fraction of the X-rays from the corona irradiate the accretion disk. These X-rays are reprocessed or reflected off the disk, and we observe the reflected emission \citet{1991}. 

The X-rays that are incident on the disk are reflected through the processes of Compton scattering and fluorescent line emission \citet{1978ApJ}. The reflected emission has an observable spectrum, which we call the X-ray reflection spectrum. This spectrum contains many different emission lines as well as the Compton hump, absorption edges, and bremsstrahlung emission. A prominent feature in the spectrum is the iron K$\alpha$ line, which is broadened by relativistic Doppler shifts thanks to the motion of matter orbiting the black hole in its accretion disk as well as general relativistic effects due to the strong gravity experienced near the black hole \citet{1989}. The closer the material is to the black hole, the stronger the relativistic effects on the disk material. This makes the reflection spectrum an effective probe of the inner accretion flow of a black hole system and allows estimates of black hole parameters such as spin to be made \citet{Reynolds_2003}. The fluorescent iron K$\alpha$ emission line is a particularly prominent feature in these reflection spectra. This prominence arises because iron is abundant in accretion disks and has a high fluorescence yield, which scales approximately as $Z^4$ where $Z$ is the atomic number for neutral or weakly ionized iron. The line is also favorable observationally because around its energy ($\sim6$ keV), Galactic absorption is negligible, as there are not many other emission or absorption lines in this energy range. Additionally, X-ray detectors also have high energy resolution around the iron line, making observations of the iron line particularly accessible. This combination of detectability and physical significance makes the iron line a useful means of measuring the characteristics of a black hole system.

The shape of the iron K$\alpha$ line profile depends on various physical parameters, including how the corona illuminates the accretion disk, the position of the innermost stable orbit ($r_{\text{ISCO}}$), the inclination of the accretion disk to the line of sight, and the geometry of the disk. Given that the shape of these line profiles depends on $r_{\text{ISCO}}$, estimations of the spin of a black hole can be made \citet{2006ApJ}. Often, estimations of both the spin and inclination of a black hole system are made by fitting the observed spectrum with a model of the reflection from the accretion disk. It is common for these models to assume a point-source corona (also known as a lamppost corona) where X-rays are emitted isotropically from some point in space above a geometrically thin, optically thick accretion disk around the black hole. These assumptions may not be accurate in describing many black hole systems, and understanding the effect of these assumptions is important for understanding biases and systematic uncertainties in measurements of fundamental parameters such as black hole spin.   

Previous investigations have been performed to understand the impact of altering the geometry of the corona \citet{Wilkins_2012, 2017} with notable alterations to the emissivity profiles from non-point-source corona geometries. Additionally, assuming an infinitesimally thin disk is an approximation for the more complex geometry of the accretion disk that we expect. The precise geometry of the accretion disks around black holes is still largely an open question, though there are four main theoretical models of accretion disks which include thin disks \citet{1973A&A}, slim disks \citet{1978MNRAS}, thick disks \citet{1976ApJ}, and advection-dominated accretion flow disks \citet{1977ApJ}. 

The Shakura-Sunyaev disk thin-disk model is arguably the most popular and well-known. In this model of accretion, the disk is geometrically thin, optically thick, and radiatively efficient. The model has been successful in a wide range of astrophysical systems, including accurately describing AGN as well as X-ray binaries. The scale height along the Shakura-Sunyaev disk increases with the Eddington ratio associated with the black hole system. The effects of the Shakura-Sunyaev disk for low Eddington ratios of $\dot{M}
 / \dot{M}_\text{Edd}= 0.1, \, 0.2,$ and $0.3$ were investigated previously by \citet{Taylor_2018} where $\dot{M}_\text{Edd}$ is the Eddington accretion rate, defined as $\dot{M}_\text{Edd} \equiv L_\text{Edd}/\eta c^2$. They found that the inclusion of disk thickness (as with the Shakura-Sunyaev disk) resulted in noticeable changes to the simulated line profiles associated with these systems, as well as an underestimation of the spin and coronal height when the disk thickness is not accounted for in the analysis. Additionally, \citet{Abdikamalov_2020} implemented the model of \citet{Taylor} and fit it to observations of GRS 1915+105. They did not find a significant difference in estimates of model parameters when using the thick disk model and the infinitesimally thin disk model. Similarly, \citet{10.1093/mnras/stac1369} used this same model and fit it to observations of MCG-06-30-15 and found that it provided a consistent spin estimate as the one inferred from fitting the infinitesimally thin disk model to the same observation. With this, it should be noted that MCG-06-30-15 is likely accreting at a low accretion rate that would fall within the regime of thin disks \citet{brenneman2025}. In addition to the Shakura-Sunyaev disk, a warped accretion disk, where the incoming stream of material is misaligned with the spin axis of the black hole, has been investigated \citet{Abarr_2021}. The line profiles from the warped disk geometry were found to be very dependent on the azimuthal angle of the observer relative to the angle of the misaligned disk. \citet{Shashank_2025} has computed the reflection spectra for super-Eddington X-ray sources, including a slim-disk and wind-reflection model. They fit these new models to the tidal disruption event Swift J1644+57 and constrained the opening angle of its outflowing funnel as well as found the associated wind to have a high velocity. In \citet{4sth-rnwv}, the state-of-the-art infinitesimally thin disk reflection models were tested against reflection spectra from general relativistic magnetohydrodynamic simulations of thin accretion disks and were found to work well for high-spin black holes, but as the simulations increased in complexity, the current reflection models started to no longer be able to accurately recover the correct spin input into the simulations. This highlights the need for reflection models that go beyond the thin disk approximation, such as those developed in this work.

In this paper, we seek to perform a comprehensive exploration of the effects of a set of different realistic disk geometries on the iron K$\alpha$ line. The contents of this paper are organized as follows. In Section \ref{sims}, we describe our ray tracing code and methods of simulating the iron K$\alpha$ line. In Section \ref{wedged_disk}, we describe the constant-aspect-ratio disk geometry and its effects on the iron K$\alpha$ line. In Section \ref{SS_disk_sec}, we detail the Shakura-Sunyaev disk and how observations of it may appear different from the flat disk. In Section \ref{expanded inner_disk_sec}, we outline how the expanded inner disk geometry appears in the iron K$\alpha$ line profile. In Section \ref{warped_sec}, we discuss the warped disk geometry and how it differs from the other disks we investigated. In Section \ref{discussion}, we discuss the effect of accretion disk geometry on estimates of the system parameters such as spin, coronal height, and inclination angle, as well as the overall implications of this work. In Section \ref{conclusion}, we summarize the results of the paper. 

\section{Simulating the Iron K$\alpha$ Line} \label{sims}

\subsection{General Relativistic Ray Tracing} \label{raytracing}

We calculate the expected iron K$\alpha$ line from different accretion disk geometries using General Relativistic ray tracing simulations, extending the code \textsc{CudaKerr} \citet{Wilkins_2012, Wilkins_2016}, to account for additional accretion disk geometries. To simulate how the accretion disk would appear to an outside observer, we trace rays from a flat image plane on the sky. This is done by integrating the geodesic equations in Kerr Spacetime. 

Following the derivation from \citet{Wilkins_2012}, in Boyer-Lindquist coordinates with $c = 1$ and $GMc^{-2} = 1$, the Kerr metric for a spinning black hole takes the following form:
\begin{equation*}
\begin{aligned}
ds^2 = & \left(1 - \frac{2r}{\rho^2}\right) dt^2  + \frac{4ar \sin^2\theta}{\rho^2} dt d\phi - \frac{\rho^2}{\Delta} dr^2 - \rho^2 d\theta^2 \\
& - \left(r^2 + a^2 + \frac{2a^2 r \sin^2\theta}{\rho^2}\right) \sin^2\theta d\phi^2
\end{aligned}
\end{equation*}
where 
\begin{align*}
    \Delta &= r^2 - 2r + a^2 \\
    \rho^2 &= r^2 + a^2\cos^2\theta
\end{align*} with units of gravitational radius, $r_\mathrm{g} = GMc^{-2}$. We are concerned with the paths of photons through spacetime near a spinning black hole. The paths of such photons are described via the null geodesics given by

\begin{equation}
    \label{eq:null_geodesic}
    g_{\mu\nu} {\dot{x}^\mu} {\dot{x}^\nu} = 0
\end{equation}

and satisfying the geodesic equation 

\begin{equation}
    \label{eq:geo}
    {\ddot{x}^\mu} + \Gamma^\mu_{\rho\sigma} {\dot{x}^\rho} \dot{x}^\sigma = 0 
\end{equation} where the metric tensor, $g_{\mu\nu}$, defines the line element $ds^2 = g_{\mu\nu} {\dot{x}^\mu} {\dot{x}^\nu}$ and determines the Christoffel symbols, $\Gamma^\mu_{\rho\sigma}$. 

By solving Equation \ref{eq:geo} with \ref{eq:null_geodesic}, we get the equations of motion of these photons through spacetime:
\begin{equation}\label{eq3}
    \dot{t}=\frac{\left[\left(r^2+a^2 \cos ^2 \theta\right)\left(r^2+a^2\right)+2 a^2 r \sin ^2 \theta\right] k-2 a r h}{r^2\left(1+\frac{a^2 \cos ^2 \theta}{r^2}-\frac{2}{r}\right)\left(r^2+a^2\right)+2 a^2 r \sin ^2 \theta}
\end{equation}
\begin{equation}\label{eq4}
    \dot{\varphi}=\frac{2 a r k \sin ^2 \theta+\left(r^2+a^2 \cos ^2 \theta-2 r\right) h}{\left(r^2+a^2\right)\left(r^2+a^2 \cos ^2 \theta-2 r\right) \sin ^2 \theta+2 a^2 r \sin ^4 \theta}
\end{equation}
\begin{equation}\label{eq5}
    \dot{\theta^2}=\frac{Q+(k a \cos \theta-h \cot \theta)(k a \cos \theta+h \cot \theta)}{\rho^4}
\end{equation}
\begin{equation}\label{eq6}
    \dot{r}^2=\frac{\Delta}{\rho^2}\left[k \dot{t}-h \dot{\varphi}-\rho^2 \dot{\theta}^2\right]
\end{equation}
where $k$, $h$, and $Q$ are conserved quantities unique to each geodesic. Additionally, the dots in these expressions signify derivatives with respect to an affine parameter that traces the position along the ray, $\sigma$.

We initialize these photons in logarithmic intervals in the field of view of the observer. The motivation for this is to maximize the number of photons that hit the inner region of the accretion disk, where the geometry and relativistic effects vary most rapidly. Each of these photons is propagated backward in time with a varying step of the affine parameter. We implement an adaptive step size; the nature of the step taken depends on the geometry of the accretion disk being simulated, which will be detailed in later sections. Generally, the step of the affine parameter is defined by 
\begin{equation}
    \label{eq:step}
    d\sigma = \frac{|\frac{r-r_H}{\dot{r}}|}{\tau}
\end{equation}
where $r$ is the distance from the event horizon, $r_H=1-\sqrt{(1-a)(1+a)}$, and $\dot{r}$ is the velocity and $\tau$ is a baseline precision parameter. Steps for each of the coordinates, $r$, $\theta$, and $\phi$, are calculated, and the smallest amongst them is used as the next step in the integration. These rays are traced until they either hit the accretion disk or miss the disk and exceed the limit of integration steps specified. Rays are said to hit the accretion disk when their corresponding stopping condition is no longer true, i.e., for a flat disk, the stopping condition is $\theta \ge \frac{\pi}{2}$. Across all the accretion disk geometries that are explored, we take the outer disk radius associated with each system to be $r_\text{disk} = 200\, r_\mathrm{g}$.

\subsection{Point-Source Corona}

In these simulations, we follow the same initialization of the point-source corona as in \citet{Wilkins_2012}. The point-source corona (also referred to as a lamppost corona) is stationary and centered on the spin axis of the black hole located at a height, $h_c$, above the accretion disk. We choose this as a simplifying assumption to explore the effects of the disk geometry by reducing the number of coronal parameters. Since the reverberation time lags measured in these systems are short, indicating the corona is compact and confined to a small region of space close to the black hole, this point-source assumption is valid \citet{2016MNRAS.462..511K, Marco_2016}. This lamppost corona is an isotropic source that emits equal power into an equal solid angle in its rest frame. An element of solid angle is given by
\begin{equation}
    \mathrm{d}\Omega' = \mathrm{d}(\cos{\beta})d\alpha
\end{equation} where $\beta$ and $\alpha$ are the polar angles. By the principle of equivalence, in the frame of the corona, we can work in locally inertial coordinates where the laws of special relativity hold, meaning we are in flat spacetime. In this frame, the rays from the corona with energy $E$ are initialized at equal intervals in solid angle and have four-momentum
\begin{equation}
    \vec{p} = (E, E\sin{\beta}\sin{\alpha}, E\sin{\beta}\cos{\alpha}, E\cos{\beta}).
\end{equation} Without loss of generality, we set $E=1$, which is equivalent to eliminating $k$ from the geodesic equations. These rays are then transformed to the global coordinate system, and the source is given its height above the accretion disk, $h_c$. By inverting the equations of motion from equations \ref{eq3}-\ref{eq6}, the constants of motion are found, and the rays can be propagated forward in time. 

\subsection{Emissivity Profiles}

In addition to tracing rays from the observer to the disk, we run a separate ray trace from the corona to the disk to get the emissivity profile. Emissivity is the variation of reflected flux (power per area) as a function of radius on the disk. By integrating the paths of photons forward in time from the corona to the accretion disk, we recover the emissivity profile, which is determined by the illumination profile of the disk by the corona.

Following \citet{Wilkins_2012}, the emissivity profile is found by counting the number of rays incident in each radial bin along the accretion disk and then dividing by the area of this disk. That is, the emissivity is given by 
\begin{equation}
    \epsilon(r) = \frac{g^2 C(r, \mathrm{d}r)}{A(r, \mathrm{d}r)}
\end{equation} where $C(r, \mathrm{d}r)$ is the number of photons in a given radial bin at radius $r$ and with a width $\mathrm{d}r$, $g \equiv \frac{E_{0}}{E_{\text{em}}}$ where $E_{\text{0}}$ is the energy of the photon viewed from an observer at the accretion disk and $E_{em}$ is the energy of the photon emitted from the corona, and $A(r, \mathrm{d}r)$ is the proper area. Both the photon arrival times and photon energy contribute a factor $g$. For more information about emissivity profiles and their behavior concerning infinitesimally thin accretion disks, see \citet{Wilkins_2012}. 

\subsection{Line Profiles}

Line profiles associated with different accretion disk systems are calculated via:
\begin{equation}\label{line_pro}
    \Phi(g) = \int\limits_{0}^{2\pi} \int_{r,\text{disk}} \int_{g'} \mathcal{E}(r, \phi)  g'^{3} \delta(g - g') T(r, \phi, g) r dg' dr d\phi
\end{equation} where $\mathcal{E}(r, \phi)$ is the emissivity as a function of radius, $r$, and azimuthal angle, $\phi$, and $g$ is the energy shift defined as before. Additionally, $T(r, \phi, g)$ is the transfer function, and it encodes the value of $g$ that is observed from different positions on the disk. $T(r, \phi, g)$ is computed from the ray tracing simulations detailed in section \ref{raytracing}, see e.g. \citet{1975ApJ...202..788C, 10.1093/mnras/staf1770}. Additionally, note the inclusion of the energy shift over the accretion disk given by $g$. By Liouville's Theorem, the specific intensity goes as $g^3$ \citet{Luminet}. We normalize these reflection spectra so that the integral of the line profile is equal to $1$. This choice of normalization facilitates comparisons between the effects of the illumination of different disk geometries for each of the explored geometries. 

\section{Constant-Aspect-Ratio Disk}\label{wedged_disk} 
\begin{figure}[b] 
    \centering
    \includegraphics[width=\linewidth]{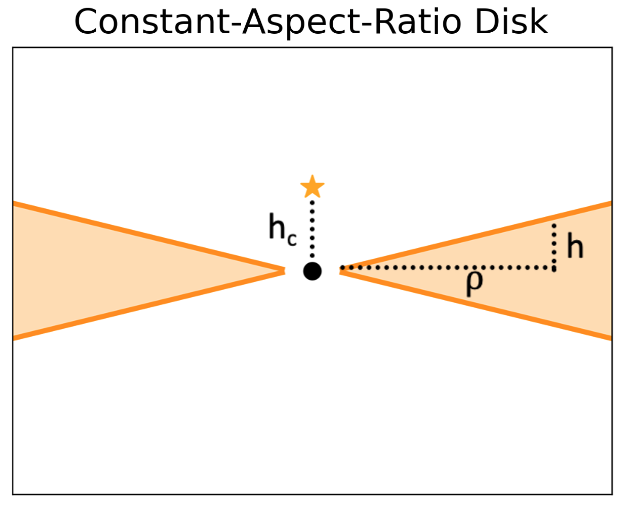}
    \caption{Cross-section of the constant-aspect-ratio accretion disk geometry viewed edge-on ($\theta = \frac{\pi}{2}$ and $\phi = 0$). The black hole is shown at the center of the disk geometry, and $h_c$ is the height of the lamppost corona above the disk. Note that this plot has no associated units and is meant to give an understanding of the basic configuration of the system.}
    \label{fig:wedged}
\end{figure}

We will now explore different accretion disk geometries, starting with the constant-aspect-ratio disk. The constant-aspect-ratio accretion disk geometry is defined by having a constant-aspect-ratio, meaning that $\frac{h}{\rho} = \text{constant}$ where $h$ is the scale height at some radius, $\rho$, along the equatorial plane. This disk geometry comes from early accretion disk theory \citet{1981ARA&A..19..137P} and is a versatile model that can be compared to a wide range of complex disk geometries in the context of reflection. Additionally, this constant-aspect-ratio disk geometry can be used as an approximation for geometrically thick flows expected in systems with super-Eddington accretion, e.g. \citet{Thomsen_2019, Dai_2018}. We assume that the material in the disk at each cylindrical radius orbits about the black hole spin axis. A schematic of the constant-aspect-ratio disk geometry is shown in Figure \ref{fig:wedged}. We investigate aspect ratios of $\frac{h}{\rho} = 0.1, 0.5,$ and $1$ to explore a range of disk thicknesses. 

To realize this disk geometry in our ray tracing simulations, we integrate the geodesic equations until the ray intersects the coordinate surface corresponding to the disk. At which point we record its location, travel time, and redshift. In the case of the constant-aspect-ratio disk, rays are stopped when their positions satisfy the condition:
\begin{equation}
    \theta \geq \frac{\pi}{2} - \arctan{\frac{h}{\rho}}.
\end{equation} We keep the variable step condition the same as described in Section \ref{raytracing}. 

\begin{figure}[htbp] 
    \centering
    \includegraphics[width=\linewidth]{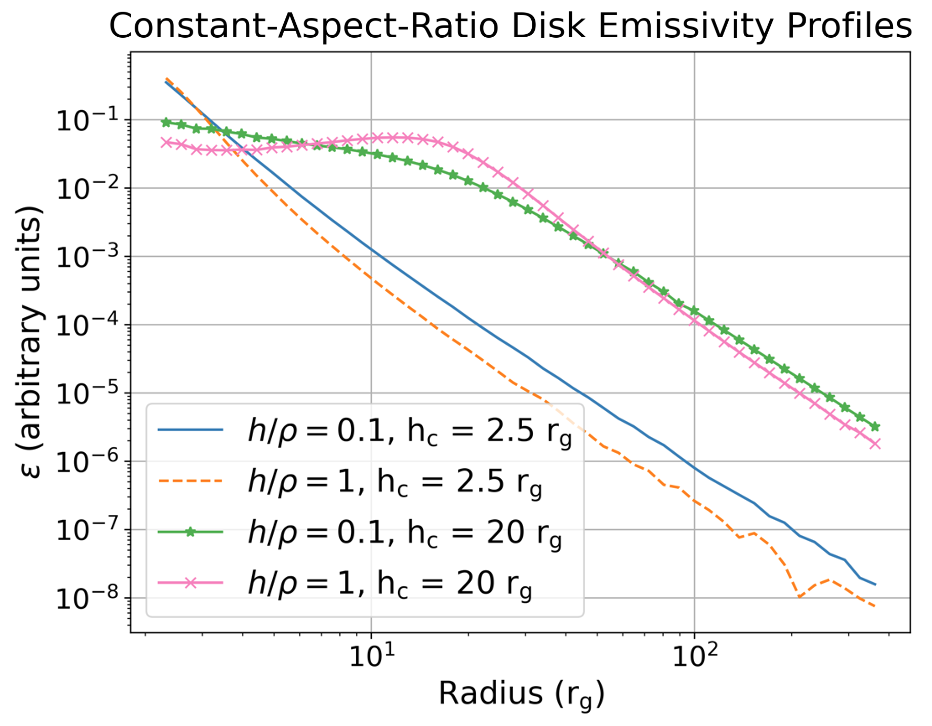}
    \caption{Emissivity profiles associated with the constant-aspect-ratio disk with $\frac{h}{\rho} = 0.1$ and $\frac{h}{\rho} = 1$ with minimum and maximum coronal heights explored ($h_c=2.5\,r_\mathrm{g}$ and $h_c=20\,r_\mathrm{g}$).}
    \label{wedge_emis}
\end{figure}

The emissivity profiles are calculated for four values of coronal height, $h_c = 2.5\, r_\mathrm{g}, \, 5 \, r_\mathrm{g},\, 10\, r_\mathrm{g},$ and $ 20\, r_\mathrm{g}$. We investigate this range of coronal heights to span the range commonly inferred for Seyfert AGN as in \citet{CACKETT2021}. Emissivity profiles associated with this geometry for different aspect ratios are found in Figure \ref{wedge_emis}. 

With a larger aspect ratio, the behavior of the emissivity profile deviates from that of a flat disk. In Figure \ref{wedge_emis}, we see the emissivity profiles associated with a disk with aspect ratio, $\frac{h}{\rho} = 0.1$ and $\frac{h}{\rho} = 1$. We do not plot the emissivity profiles for the flat disk as they are very similar to those of the $\frac{h}{\rho} = 0.1$ disk. For the lowest coronal height of $h_c = 2.5 \, r_\mathrm{g}$, we see that an increased aspect ratio causes an increase in rays that hit the inner accretion flow. Additionally, for the larger coronal height $20 \, r_\mathrm{g}$, there is an increase in the number of rays that hit the disk near $\rho \approx h_c$ in the disk with $\frac{h}{\rho} = 1$ versus the thinner disk with $\frac{h}{\rho} = 0.1$. This enhancement in the rays in this region is the byproduct of the geometry of the disk. The extreme thickness of the disk is such that the disk is intercepting rays near $\rho \approx h_c$ that would have otherwise been collected in the very inner radii of a flat-disk system. 

\begin{figure*}
\includegraphics[width=1\textwidth]{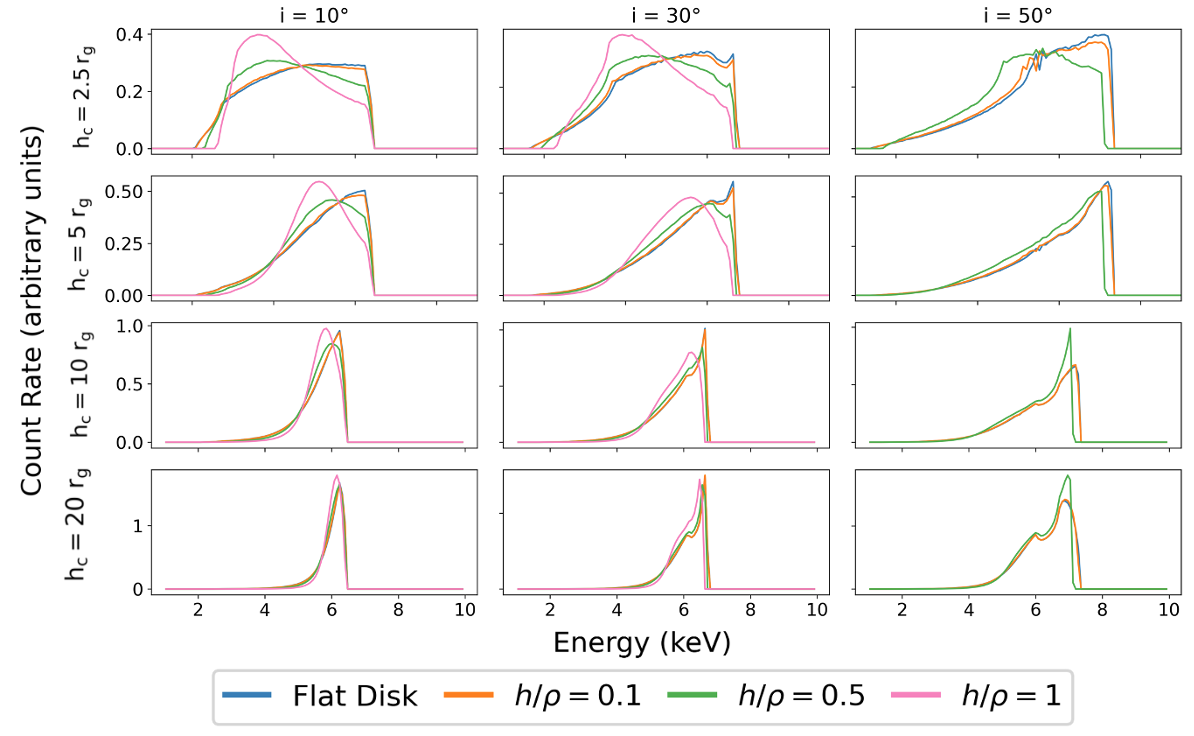}
    \caption{Line profiles from the constant-aspect-ratio accretion disk. The top row shows the profiles associated with a coronal height of $2.5 \, r_\mathrm{g}$, the middle rows are the same but for coronal heights of $5 \, r_\mathrm{g}$ and $10 \, r_\mathrm{g}$, and the last is for $20 \, r_\mathrm{g}$. The columns correspond to different inclination angles of observation, $i$. The blue line corresponds to the flat disk, the orange line corresponds to the disk with $\frac{h}{\rho} = 0.1$, the green line corresponds to $\frac{h}{\rho} = 0.5$, and the pink line corresponds to $\frac{h}{\rho} = 1$.}
    \label{wedge_lines}
\end{figure*}

\begin{figure*}
\includegraphics[width=1\textwidth]{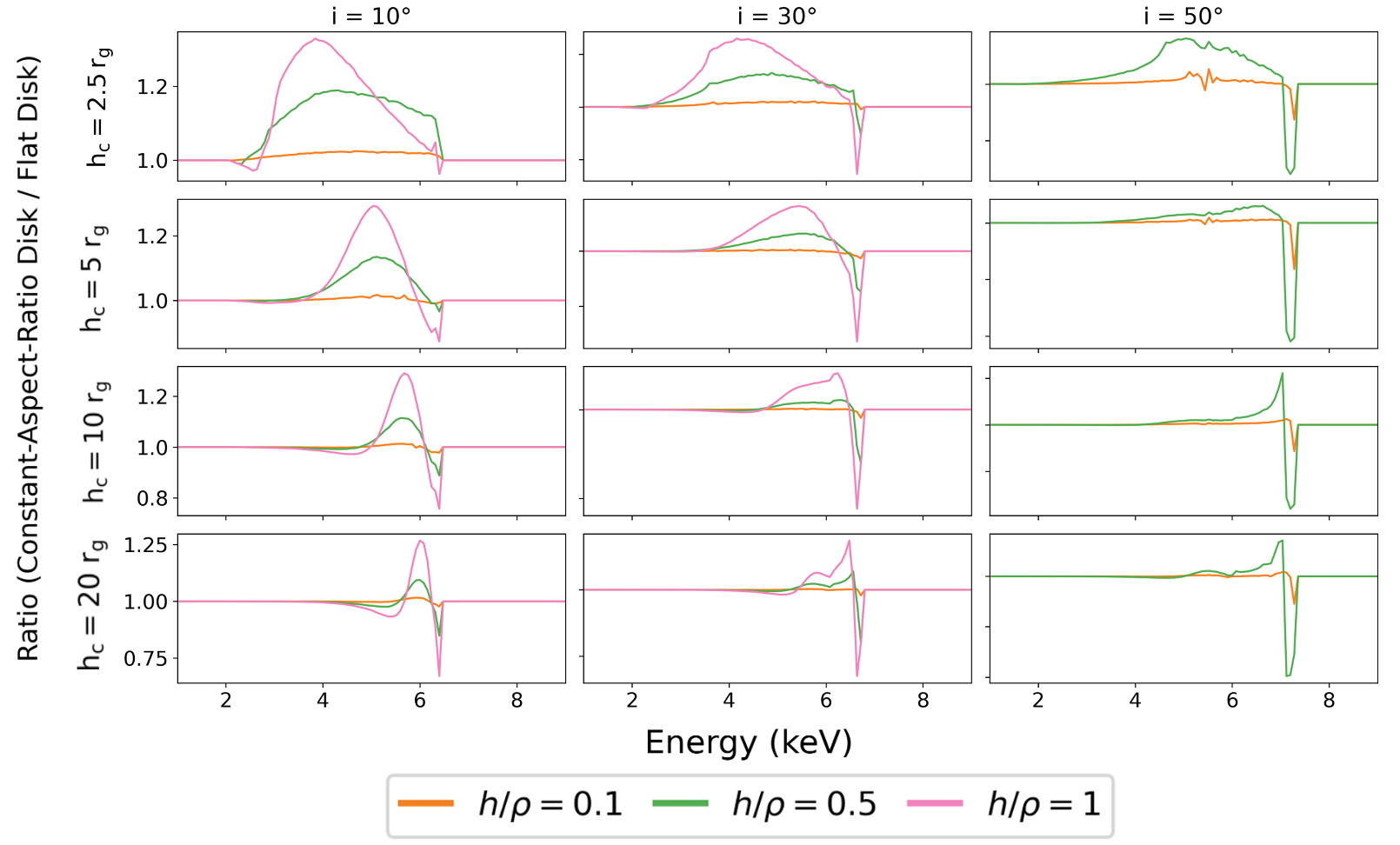}
    \caption{Ratio plots of the constant-aspect-ratio disk line profiles and those from the flat disk. Each line is the division of the constant-aspect-ratio disk line by the flat disk line with both having the same coronal height and observation inclination, $i$.}
    \label{wedge_ratios}
\end{figure*}

In Figure \ref{wedge_lines}, we have the line profiles associated with the constant-aspect-ratio disk for various coronal heights, observation inclination angles, and disk thicknesses, as well as the line profiles from the flat-disk geometry. Additionally, in Figure \ref{wedge_ratios}, we have the associated plots of the ratio of the line profiles from the constant-aspect-ratio disk divided by the line profiles from the equivalent flat disk. When we calculate the ratio of the spectra, we add in the primary continuum component, which is approximated by a power law with a photon index of $2$. These plots are created by adding a power law of index $2$ to the line profiles seen in Figure \ref{wedge_lines} and then dividing them by the power law plus the line profiles from the corresponding flat disk. 

\begin{figure*}
\includegraphics[width=1\textwidth]{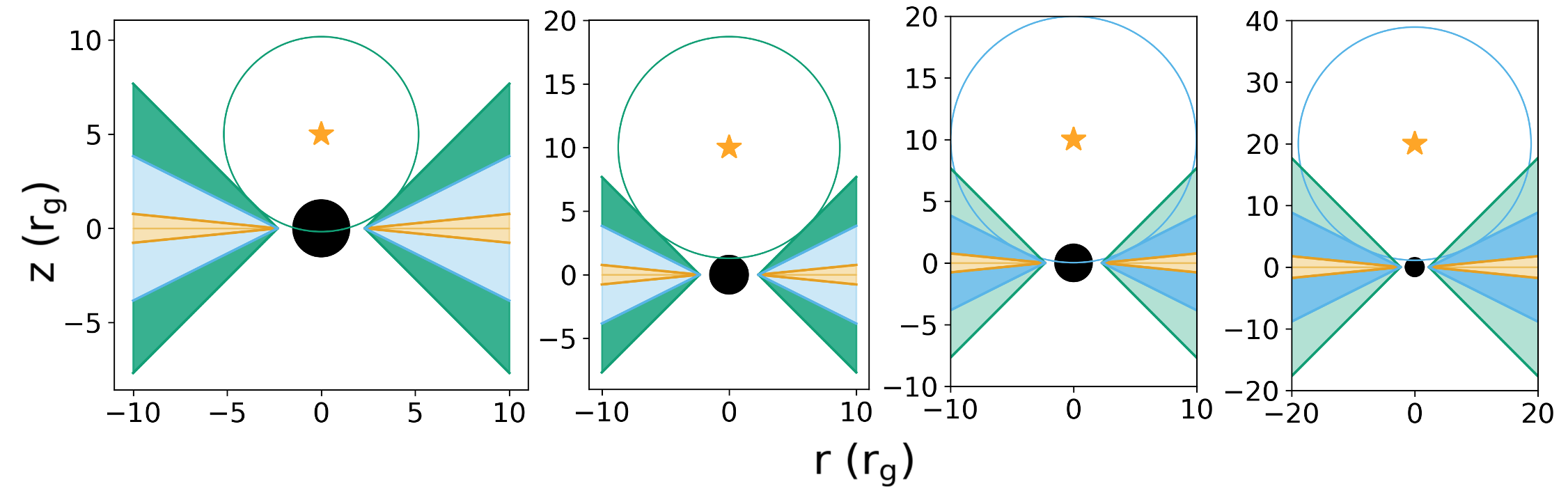}
    \caption{Schematic of the constant-aspect-ratio disk geometries with a lamppost corona emitting constant emissivity from a boundary (circle in each panel). The green disk has $\frac{h}{\rho} = 1$, the blue disk has $\frac{h}{\rho} = 0.5$, and the orange disk has $\frac{h}{\rho} = 0.1$. In the leftmost panel, with $h_c = 5 \, r_\mathrm{g}$, the region of constant emissivity makes first contact with the inner radii of each constant-aspect-ratio disk. In the second panel from the left, $h_c = 10 \, r_\mathrm{g}$, and we can see that this region of constant emissivity intersects the disk at larger radii for $\frac{h}{\rho} = 0.1$. In the third panel from the left, we see that an equivalent region of constant emissivity still reaches the inner radii of the disks with $\frac{h}{\rho} = 0.1$ and $\frac{h}{\rho} = 0.5$. In the rightmost panel, at $h_c = 20 \, r_\mathrm{g}$, this region instead intersects the disks with $\frac{h}{\rho} = 0.5$ at larger radii. The lack of illumination of the inner accretion disk in each case leads to a corresponding decrease in the number of the most highly redshifted photons in the resulting line profiles.}
    \label{wedge_beh}
\end{figure*}

Across all coronal heights, there is a reduction in the number of the most redshifted photons in the line profiles in the range of energies from $3-5$ keV. We directly see this in Figure \ref{wedge_emis} in the reduction in emissivity at the inner disk. The explanation for this effect is described in Figure \ref{wedge_beh}. In this figure, we show the surface of constant flux from the corona. These circles shown are the smallest surfaces of constant flux that intersect the constant-aspect-ratio disk. For the thicker disks with aspect ratios of $\frac{h}{\rho} = 0.5$ and $1$, the surface of constant flux is incident on radii which are considerably greater than $r_\text{ISCO}$, resulting in the reduction in the number of the most redshifted photons seen mostly in the plots with coronal heights of $10\, \text{r}_\text{g}$ and $20\, \text{r}_\text{g}$ in Figure \ref{wedge_ratios}. Additionally, we see an increase in the number of moderately redshifted rays, caused by this increase in rays that hit the constant-aspect-ratio disk at radii near $\rho \simeq h_c$. This is followed by a reduction in the number of photons near $6.4$ keV as fewer rays hit the outer radii of these constant-aspect-ratio disks. 

\subsection{Compressed Inner Disk}

In addition to the constant-aspect-ratio disk geometry, we investigate a compressed inner disk. In this geometry, the inner accretion flow consists of a flat disk extending from $r_\text{ISCO}$ to a break radius $\text{r}_\text{b}$, beyond which the disk transitions to a constant aspect ratio (or constant-aspect-ratio shape). This is an extension of the constant-aspect-ratio disk geometry, which is motivated by the prospect of having magnetic field lines associated with a jet or corona that cause compression of the inner accretion flow \citet{1994ApJ...436..599S}. We explore a set of compressed inner disks with the following parameters: $\text{r}_\text{b} = 5\, \text{r}_\text{g}, \, 10 \,\text{r}_\text{g}, \, 20 \,\text{r}_\text{g}$; $\text{h}_\text{c} = 2.5\, \,\text{r}_\text{g}, \, 5\,\text{r}_\text{g}, \, 10 \,\text{r}_\text{g}, \, 20 \,\text{r}_\text{g}$; $\frac{h}{\rho} = 0.1, \, 0.5, \, 1$. To simulate this geometry in our ray tracing code, we use the stopping condition $
\theta \ge \frac{\pi}{2}$ when $\rho < r_b$ to propagate the rays to the accretion disk. When $\rho > r_b$, we use the following condition:
\begin{equation}
    \theta \ge \Theta + \arcsin{\frac{r_b \cos{\Theta}}{r}}
\end{equation} where $\Theta \equiv \frac{\pi}{2}-\arctan{\frac{h}{\rho}}$. Similarly, we use this condition to determine the variable step in the propagation of the rays to the accretion disk. 
\begin{figure}[htbp] 
    \centering
    \includegraphics[width=\linewidth]{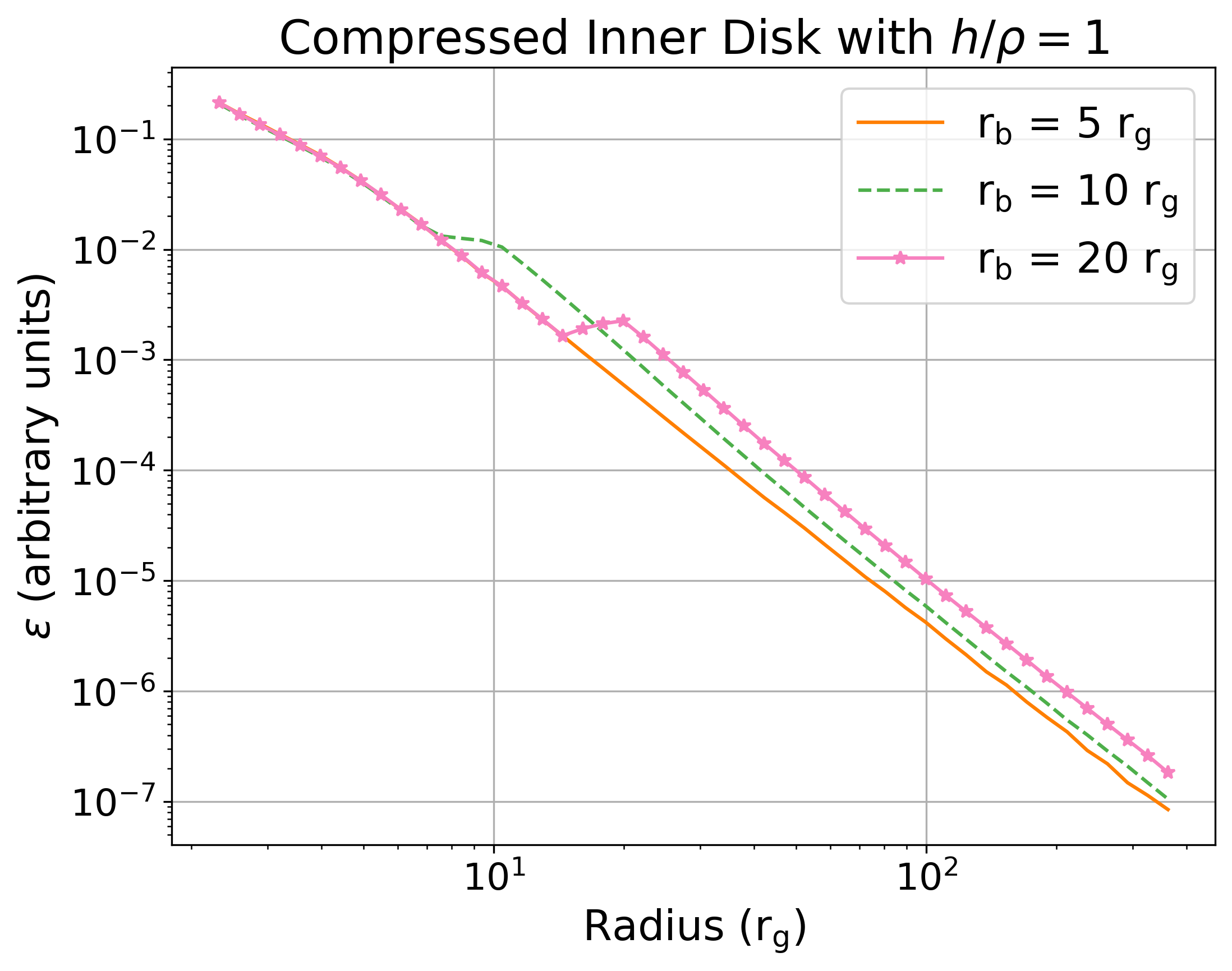}
    \caption{Emissivity profiles associated with the compressed inner accretion disk with $\frac{h}{\rho} = 1$. Each line corresponds to a disk illuminated by a corona at height $5\, \text{r}_\text{g}$, with different break radii $r_b$.}
    \label{delayed_wedge_emis}
\end{figure}

The emissivity profiles of the compressed inner disk behave similarly to the constant-aspect-ratio disk but with the inclusion of an additional feature. This is seen in Figure \ref{delayed_wedge_emis} where the associated emissivity profiles for the compressed inner disk with different values of $\text{r}_\text{b}$ are shown. From these emissivity profiles, we see that there is an increase in the number of rays that hit the accretion disk at $\rho \approx r_b$; this effect is more prevalent for larger coronal heights. This increase in emissivity at these radii is the result of the constant-aspect-ratio geometry being physically closer to the corona than a flat disk otherwise would be. 

\begin{figure*}
\includegraphics[width=1\textwidth]{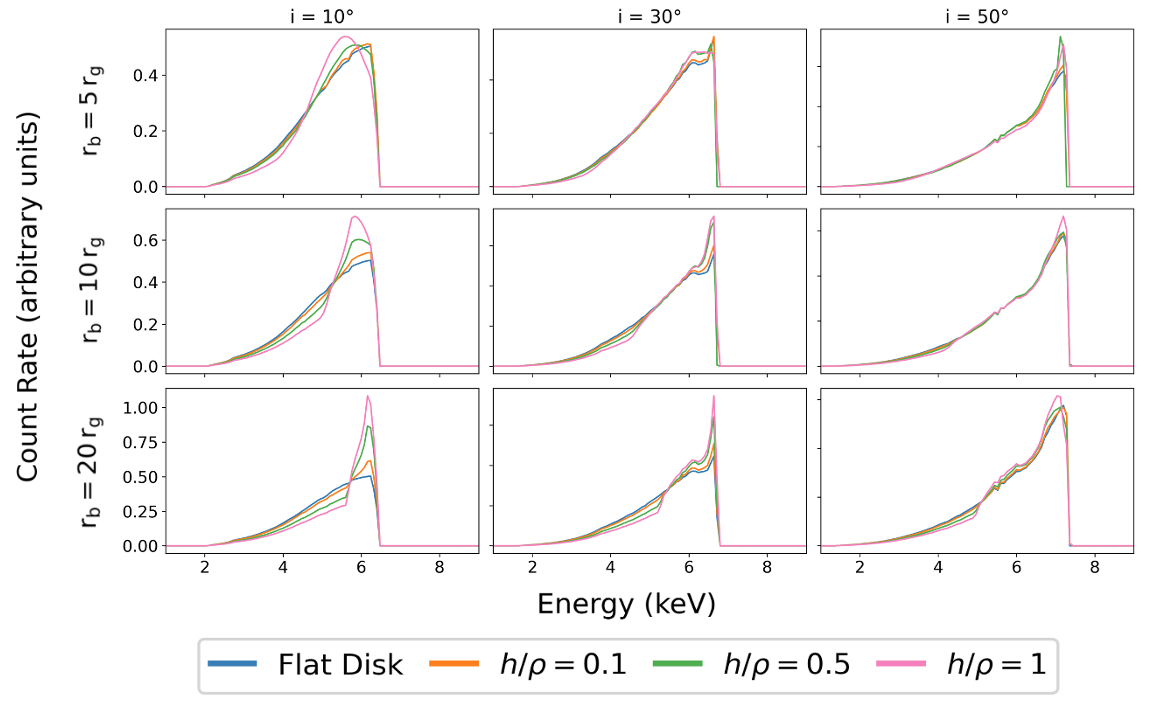}
    \caption{Line profiles from the compressed inner accretion disk with $\text{h}_\text{c}=5\,\text{r}_\text{g}$. The top row shows the profiles associated with a break radius of $\text{r}_\text{b} = 5 \, r_\mathrm{g}$, the middle row is the same but for $\text{r}_\text{b} = 10 \, r_\mathrm{g}$, and the last is for $\text{r}_\text{b} = 20 \, r_\mathrm{g}$. The columns correspond to different inclination angles of observation, $i$. A key labeling the corresponding colors for each disk is shown at the bottom of the plot.}
    \label{delayed_wedged_lines}
\end{figure*}

The line profiles of the compressed inner disk with $\text{h}_\text{c} = 5\,\text{r}_\text{g}$ are shown in Figure \ref{delayed_wedged_lines}. In these line profiles, we see there is a reduction in the number of the most redshifted photons. This reduction in the most redshifted photons is greater in this disk geometry than it was for the constant-aspect-ratio disk. The additional reduction in photons reaching the inner accretion flow arises because the inner flat disk lies physically farther from the corona than the corresponding regions of the constant-aspect-ratio disk. At the constant-aspect-ratio disk portion of the compressed inner disk, there is a large increase in the number of photons incident on the disk, producing the enhancement seen in the line profiles at$\sim 5$ keV. 

\section{The Shakura-Sunyaev Disk}\label{SS_disk_sec}
The Shakura-Sunyaev model is one of the most widely used accretion disk models, first presented by \citet{1973A&A}. For this accretion disk model, when the inner accretion flow is radiation-pressure dominated, the disk scale height depends on the black hole spin, radiative efficiency, and radial distance from the black hole. The scale height in this regime is given by:
\begin{equation} \label{SS_eq} 
H = \frac{3}{2} \frac{1}{\eta} \left( \frac{\dot{M}}{\dot{M}_{\text{Edd}}} \right) \left[ 1 - \sqrt{\frac{r_{\text{ISCO}}}{r \sin \theta}} \right]
\end{equation}
where $\eta$ is the radiative efficiency of the disk, $\dot{M} / \dot{M}_{\text{Edd}}$ is the Eddington ratio. To realize these accretion disk geometries in our simulations, we alter the stopping condition for the rays to the following:
\begin{equation} 
z \le \frac{3}{2} \frac{1}{\eta} \left( \frac{\dot{M}}{\dot{M}_{\text{Edd}}} \right) \left[ 1 - \sqrt{\frac{r_{\text{ISCO}}}{r \sin \theta}} \right]
\end{equation} where $z$ is the position where a given ray must stop along the z-axis. Additionally, the conditions for the variable steps of integration taken are the same as described by Equation \ref{eq:step} but with the value of baseline precision parameter $\tau$ increased. 

\begin{figure}[htbp] 
    \centering
    \includegraphics[width=\linewidth]{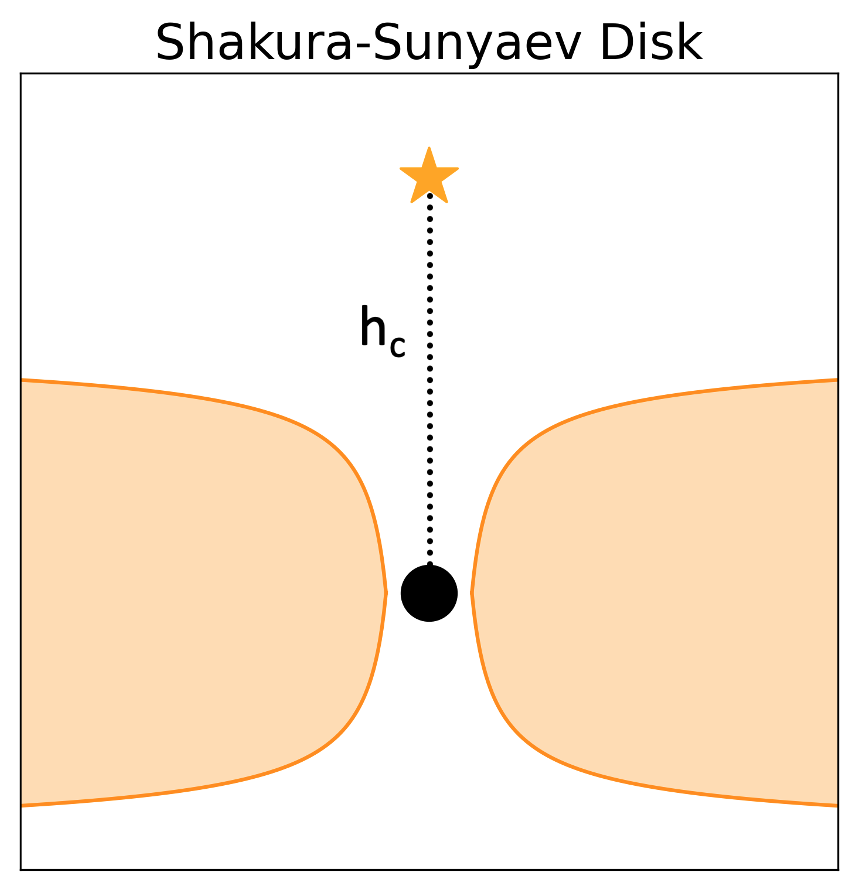}
    \caption{Cross-section of the Shakura-Sunyaev accretion disk geometry viewed edge-on, $\theta = \frac{\pi}{2}$ and $\phi = 0$. Black hole shown at center of the disk geometry and $h_c$ is the height of the lamppost corona above the disk. Note this plot has no associated units and is meant to give an understanding of the basic configuration of the system.}
    \label{SS_disk}
\end{figure}

This prescription of the Shakura-Sunayev disk was used by \citet{Taylor_2018} for low Eddington ratios given by $\dot{M} / \dot{M}_\text{Edd}= 0.1, \, 0.2,$ and $0.3$. We investigate larger Eddington ratios of $\dot{M} / \dot{M}_\text{Edd}=0.3, \, 0.7, \, 1.1,$ and $17$ with $\eta = 0.35$. A depiction of the Shakura-Sunyaev disk is shown in Figure \ref{SS_disk}. The Shakura-Sunyaev disk is usually meant to have accretion rates that are sub-Eddington; however, we extend the use of this accretion model to larger rates of accretion, as it still provides a basic approximation for these higher Eddington and super-Eddington accretion systems. The additional choice of super-Eddington disk with $\dot{M} / \dot{M}_\text{Edd}=17$ was made to approximate the super-Eddington disk model obtained from general relativistic magnetohydrodynamic simulations shown in Figure $1$ in \citet{Thomsen_2019}. With all of these Shakura-Sunyaev disks, we assume all material to have Keplerian orbits. This approximation will break down as the accretion disk becomes super Eddington. In this case, radiation pressure will start to drive radial motion of the disk surface, as is the case in \citet{Thomsen_2019, Dai_2018}. This effect would especially be important in our simulation of the Shakura-Sunyaev disk with $\dot{M} / \dot{M}_\text{Edd}=17$. 

\begin{figure}[htbp] 
    \centering
    \includegraphics[width=\linewidth]{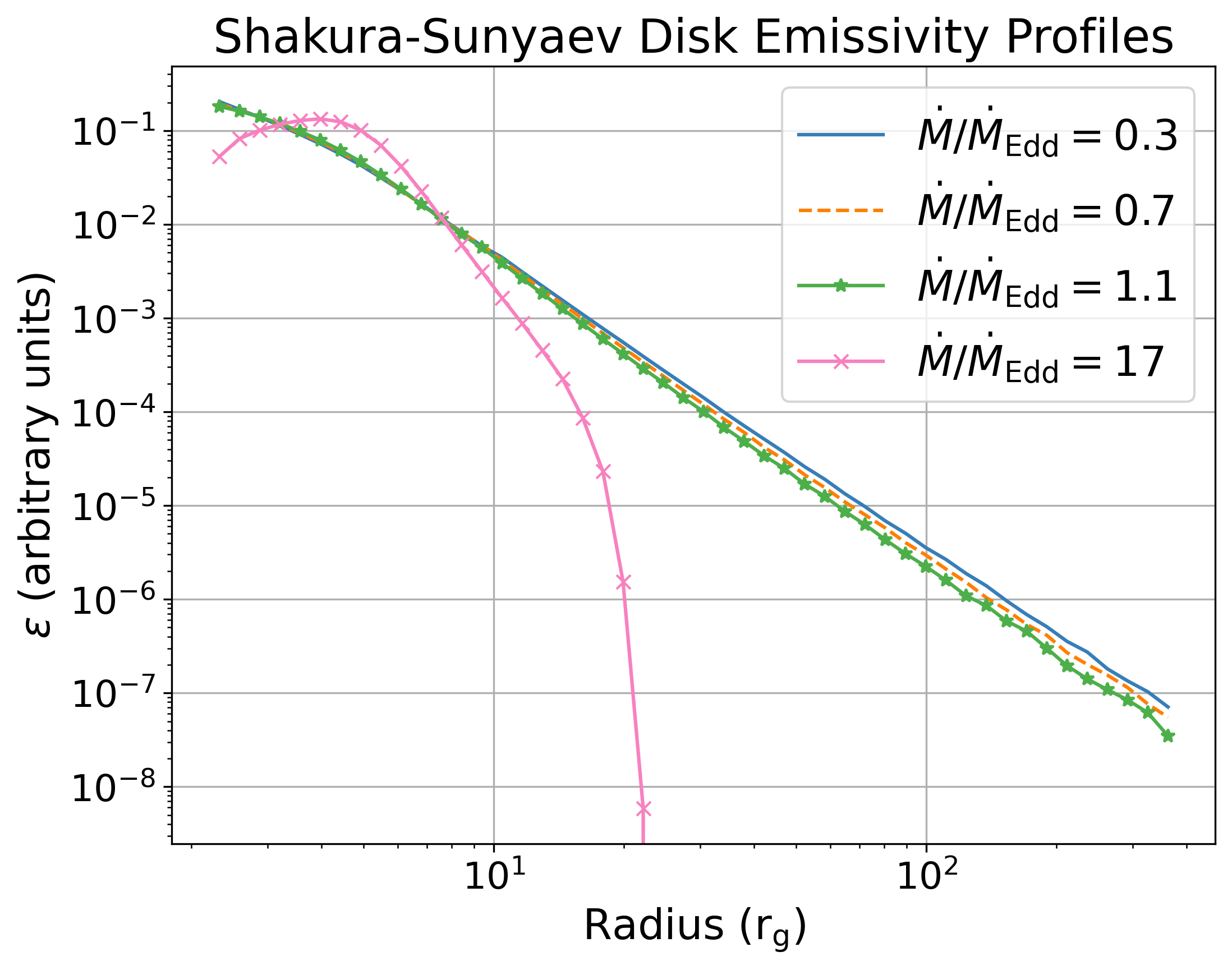}
    \caption{Emissivity profiles associated with the Shakura-Sunyaev accretion disk with a coronal height of $5\, r_\mathrm{g}$ and Eddington ratios $\dot{M} / \dot{M}_{\text{Edd}} = 0.3$, $0.7$, $1.1$, and $17$. Each line corresponds to the respective accretion disk geometry, but with a different Eddington ratio.}
    \label{ss_emissivity}
\end{figure}

The emissivity profiles associated with the Shakura-Sunyaev disk with $\dot{M} / \dot{M}_\text{Edd}=0.7$ are shown in Figure \ref{ss_emissivity}. With these emissivity profiles, we see the effect of the disk thickness. The inner accretion flow collects more rays than the flat disk as a result of the non-negligible scale height of the Shakura-Sunyaev disk. This effect is not exactly the same as with the constant-aspect-ratio disk. As one increases the thickness of the Shakura-Sunyaev disk, there is no enhancement of the rays at intermediate radii along the disk as occurs for the constant-aspect-ratio disk (shown in Figure \ref{wedge_beh}). This is largely since the scale-height of the Shakura-Sunyaev disks that we investigate are significantly smaller than the constant-aspect-ratio disks with larger aspect ratios of $\frac{h}{\rho}=0.5$ and $1$. In the case of the Shakura-Sunyaev disk with $\dot{M} / \dot{M}_\text{Edd}=17$, a similar enhancement of rays at intermediate radii is observed as with the thicker constant-aspect-ratio disks.

\begin{figure*}
    \centering
    \includegraphics[width=\linewidth]{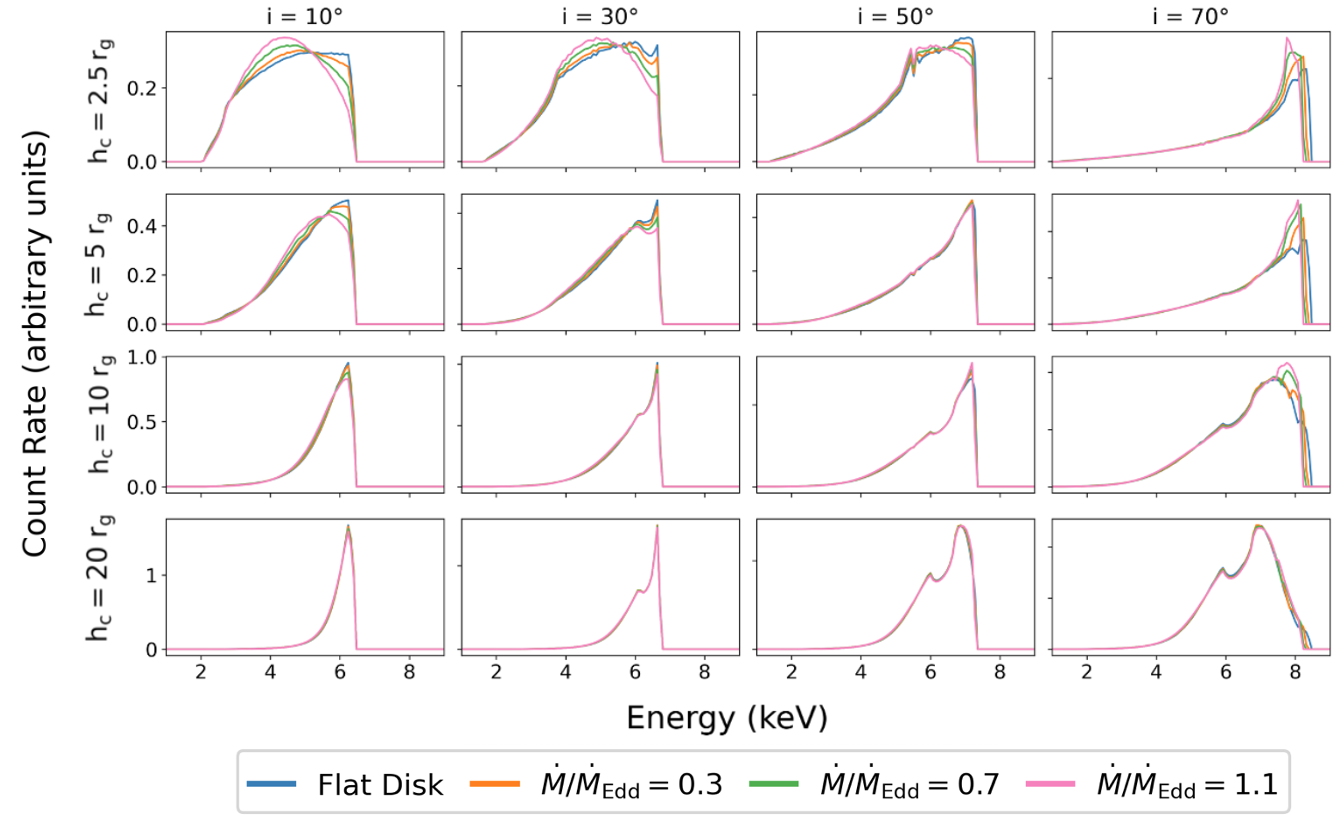}
    \caption{Line profiles from the Shakura-Sunyaev accretion disk. The top row shows the profiles associated with a coronal height of $2.5 \, r_\mathrm{g}$, the middle rows are the same but for coronal heights of $5 \, r_\mathrm{g}$ and $10 \, r_\mathrm{g}$, and the last is for $20 \, r_\mathrm{g}$. The columns correspond to different inclination angles of observation, $i$. The blue line corresponds to the flat disk, the orange line corresponds to the disk with $\dot{M} / \dot{M}_{\text{Edd}} = 0.3$, the green line corresponds to $\dot{M} / \dot{M}_{\text{Edd}} = 0.7$, and the pink line $\dot{M} / \dot{M}_{\text{Edd}} = 1.1$.}
    \label{ss_lines}
\end{figure*}

\begin{figure*}
    \centering
    \includegraphics[width=\linewidth]{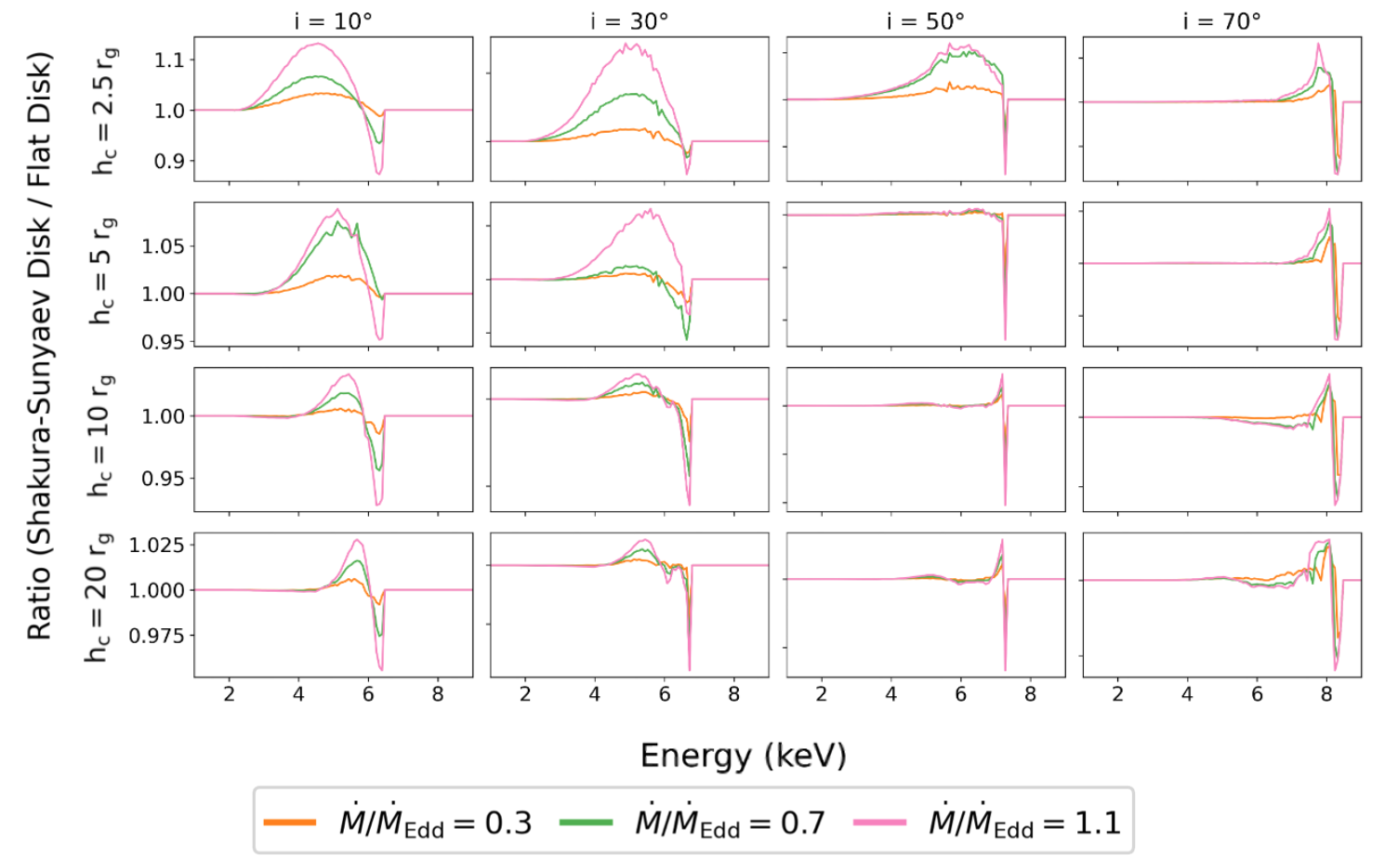}
    \caption{Ratio plots of the Shakura-Sunyaev disk line profiles and those from the flat disk. Each line is the division of the Shakura-Sunyaev disk line by the flat disk line at the same coronal height and observation inclination, $i$.}
    \label{ss_ratios}
\end{figure*}

In addition to the emissivity profiles, we have the line profiles associated with each of these different Shakura-Sunayev disks shown in Figure \ref{ss_lines} as well as the accompanying ratio plots of the lines from the Shakura-Sunyaev disk plus a power law of index $2$ divided by the flat disk plus a power law of index $2$ in Figure \ref{ss_ratios}.

The increase in the scale height of the accretion disk promotes changes in the reflection spectra. Most notably, as with the constant-aspect-ratio disk at low coronal height, the Shakura-Sunyaev disk creates an enhancement in the number of highly redshifted photons as the thickness of the disk is increased. In Figure \ref{ss_ratios}, this trend is seen in each of the panels (roughly between the energies of $2.5-5$ keV). For many of the inclination angles of observation, an increase in the number of the most redshifted photons and a decrease in the counts of moderately redshifted and blueshifted photons is observed.  

At lower inclination angles of $10^\circ$, we see an increase in the redshifted photons below $\sim 5$ keV and a decrease in the rays that are moderately redshifted and blueshifted near $6.4$ keV. With increasing coronal height, these effects become less prominent as the disk thickness becomes less significant. For an inclination of $30^\circ$, there is an increase in the most redshifted photons and in the number of moderately redshifted as well as a shifting of the blue peak of the line profiles. This slight shifting of the peak between lines creates a large dip in the ratio plots as seen near $6.4$ keV in the second column of Figure \ref{ss_ratios}. The shift in the blue peak results from the reduction in the number of blueshifted rays associated with the increased contribution from more strongly redshifted rays. For higher observation inclinations of $50^\circ$ and $70^\circ$, there is an enhancement in the number of rays below $\sim 5$ keV and a reduction in the number of photons with energies between $5-7$ keV and an increase in the most blueshifted rays above $\sim 7$ keV. 

The increase in the most redshifted photons across all line profiles is caused by the nonnegligible scale height of this disk. That is, additional photons from the corona hit the inner accretion flow of these Shakura-Sunyaev disks. The thickness of the disk causes the outer regions to intercept fewer rays because a larger fraction of rays hit the inner accretion disk. These photons, which hit the outer portions of the disk, have associated energies near $6.4$ keV. Namely, the outer portions of the disk experience less gravitational redshift from the central black hole and therefore experience less extreme energy shifting. 

\subsection{Shakura-Sunyaev Disk with $\dot{M} / \dot{M}_{\text{Edd}} = 17$}

\begin{figure*}
    \centering
    \includegraphics[width=\linewidth]{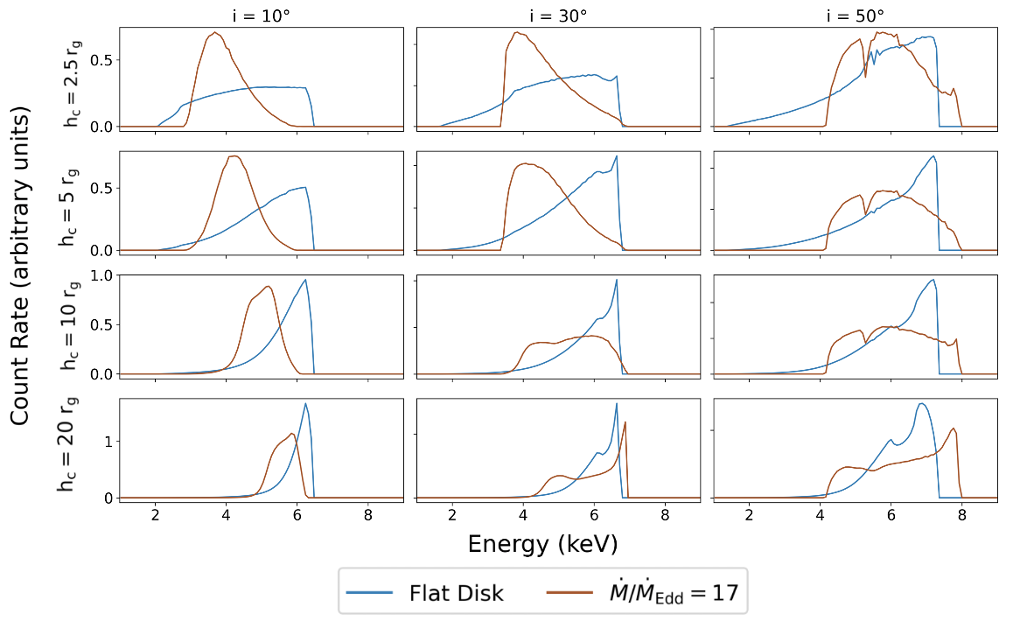}
    \caption{Line profiles from the Shakura-Sunyaev accretion disk with super-Eddington rate of $\dot{M} / \dot{M}_{\text{Edd}} = 17$. The top row shows the profiles associated with a coronal height in decreasing order. The columns correspond to different inclination angles of observation, $i$. The blue line corresponds to the flat disk and the brown line corresponds to the disk with $\dot{M} / \dot{M}_{\text{Edd}} = 17$.}
    \label{ss_lines_17}
\end{figure*}

We investigate a Shakura-Sunyaev disk in an extreme accretion scenario in which its Eddington ratio is $17$. We choose this Eddington rate as it approximates the geometry of the super-Eddington systems such as those in \citet{Thomsen_2019}. The line profiles from this accretion scenario are shown in Figure \ref{ss_lines_17}. These profiles are shown only for low inclination angles, as the majority of rays are blocked by the accretion disk at higher inclinations. This geometry produces extreme redshifting because most of the rays emitted from the corona hit the inner regions of the accretion flow. In this extreme configuration, the corona resides in a ‘funnel’ of the accretion disk, making it difficult for photons to reach the outer radii of the disk. At an inclination of $50^\circ$, the effect of this extreme geometry becomes even more prominent as the number of photons incident on the inner accretion flow gets shielded by the accretion disk, causing a large enhancement in the red wing of the line profiles. This extreme disk geometry would be detectable with XRISM. In these simulations, we assume Keplerian orbits. However, in general-relativistic magnetohydrodynamic (GRMHD) simulations of super-Eddington systems, such as in \citet{Thomsen_2019}, the material in the disk has nonzero radial velocities caused by radiation-pressure-driven winds. These radial velocities cause additional blueshifting as the emitting material moves towards the observer. 

\section{The Expanded Inner Disk} \label{expanded inner_disk_sec}
We consider an accretion disk geometry that consists of an inner accretion disk that is radiative-pressure-dominated, while the outer radii consist of a flat disk. This geometry approximates a system where extreme radiation pressure around the inner accretion flow causes the accretion disk to become inflated. For example, this disk could resemble a tidal disruption event super-Eddington disk with a supermassive black hole. A disk of this type can be seen in the general relativistic magnetohydrodynamic simulations (GRMHD) performed by \citet{Dai_2018} (see Figure $1$ in their paper for a qualitative image of their results). In this analysis, this geometry is approximated by the expanded inner disk geometry to see how it would appear in the iron K$\alpha$ line. 

\begin{figure}[htbp] 
    \centering
    \includegraphics[width=\linewidth]{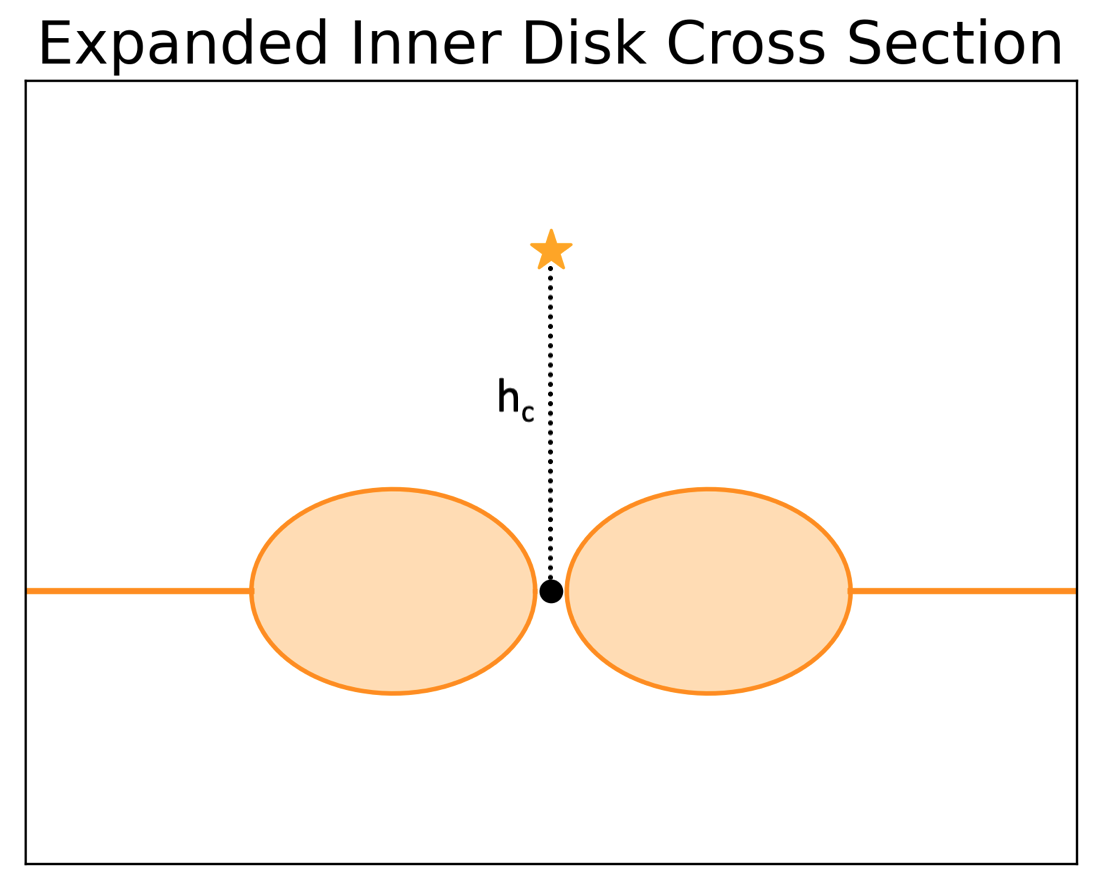}
    \caption{Cross-section of the expanded inner accretion disk geometry viewed edge-on, $\theta = \frac{\pi}{2}$ and $\phi = 0$. Black hole shown at center of the disk geometry and $h_c$ is the height of the lamppost corona above the disk. Note this plot has no associated units and is meant to give an understanding of the basic configuration of the system.}
    \label{ellipse_disk}
\end{figure}

To create the expanded inner disk geometry, we use the Shakura-Sunyaev disk solution to find a radius at which an accretion disk transitions from being radiative-pressure dominated to gas-pressure dominated. This radius is used as the length of the major axis of the expanded inner disk. We take the semi-minor axis of the expanded inner cross-section to be the scale height from the Shakura-Sunyaev solution from Equation \ref{SS_eq}. Outside of the expanded inner part of the disk geometry, we attach a flat disk to the outer radii. This is done to ensure all of our disk geometries have the same farthest outer radius, as well as ensure that the line profiles from this geometry would have a well-developed blue wing. Additionally, this outer flat disk approximates the debris stream seen in the GRMHD simulations as referenced above. In Figure \ref{ellipse_disk}, we have a schematic of this accretion disk geometry. 

To realize this disk geometry, we use the following stopping condition in the region from $r_\text{ISCO} < r < r_\text{expanded}$  ($r_\text{expanded}$ is the radius at which the disk no longer is an expanded inner disk but becomes a flat disk): 
\begin{equation}
    z \le \sqrt{b^2(1-(n-m)^{2}a^{-2})}
\end{equation} where $a$ is the semimajor axis of the ellipse, $b$ is the semiminor axis of the ellipse, $n = \sqrt{x^2+y^2}$ in Cartesian coordinates, and $m = r_\text{ISCO}+a$. Then, in the region from $r_\text{expanded} < r < r_\text{outer}$, the condition is the simply $\theta \geq \frac{\pi}{2}$ for the outer flat disk. The steps of integration are the same as those defined in Equation \ref{eq:step}. 

For the expanded inner disk geometry, we simulate a suite of line profiles, varying the observation, scale height of the semi-minor axis of the ellipse geometry, and height of the corona above the disk. We use the same observation inclination angles as before and use Equation \ref{SS_eq} to find the values of the semi-minor axis with the Eddington ratios $\dot{M} / \dot{M}_{\text{Edd}} = 0.3, \, 0.7$, and $1.1$.

\begin{figure}[htbp] 
    \centering
    \includegraphics[width=\linewidth]{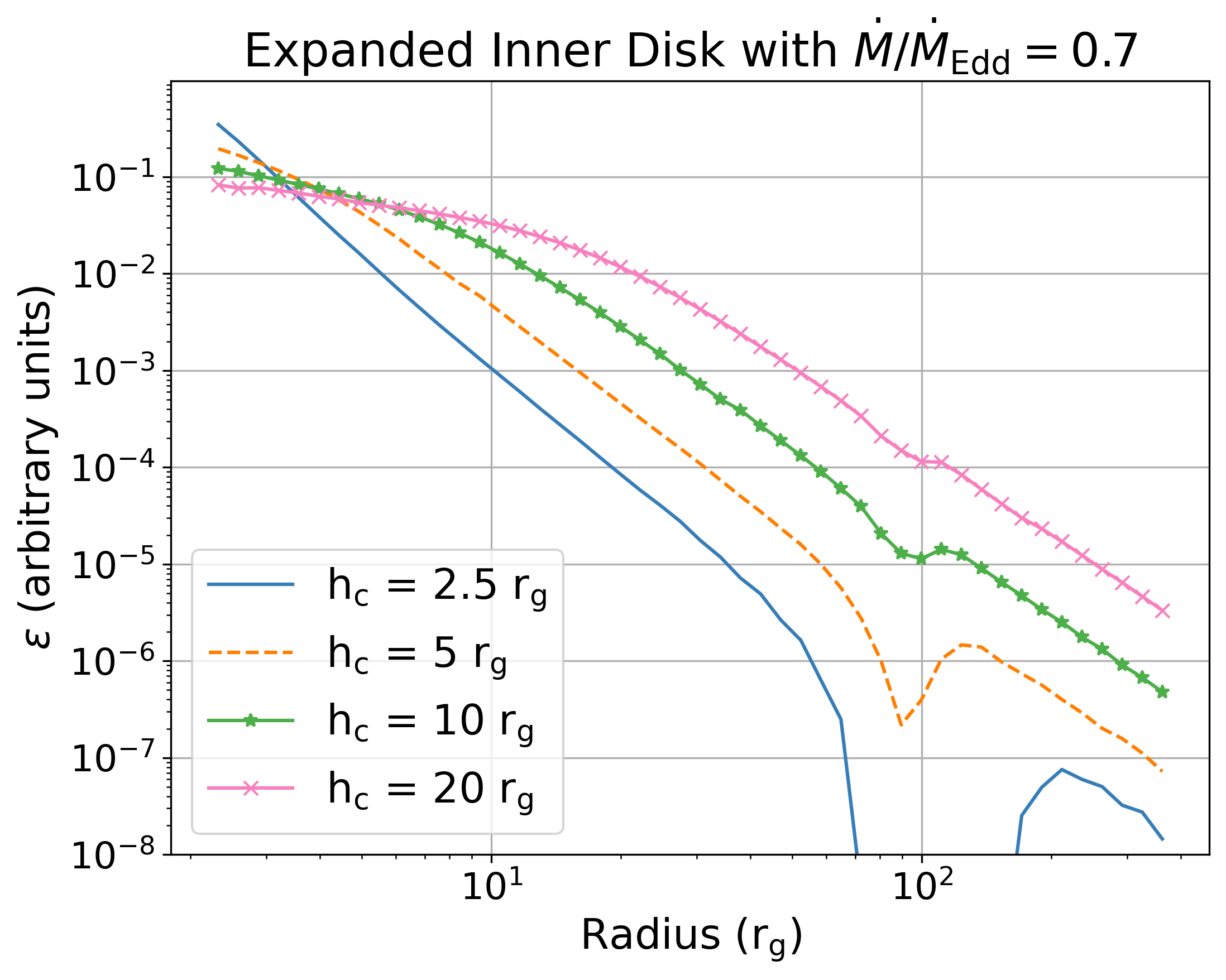}
    \caption{Emissivity profiles associated with the expanded inner accretion disk with Eddington ratio $\dot{M} / \dot{M}_{\text{Edd}} = 0.7$. Each line corresponds to the same accretion disk geometry but with a different coronal height, $h_c$, above the disk.}
    \label{ellipse_emis}
\end{figure}

\begin{figure*}
    \centering
    \includegraphics[width=\linewidth]{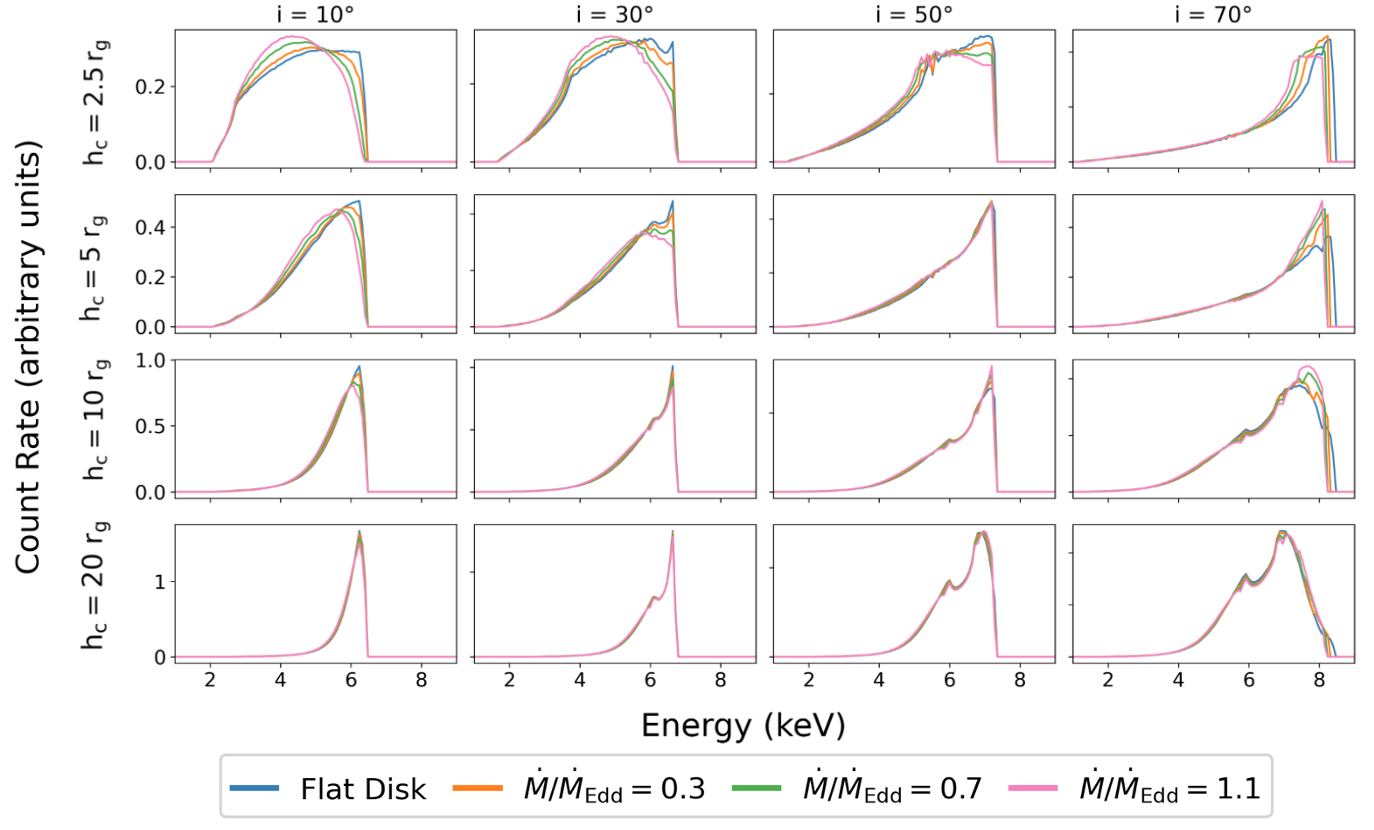}
    \caption{Line profiles from the expanded inner accretion disk. The top row shows the profiles associated with a coronal height of $2.5 \, r_\mathrm{g}$, the middle rows are the same but for coronal heights of $5 \, r_\mathrm{g}$ and $10 \, r_\mathrm{g}$, and the last is for $20 \, r_\mathrm{g}$. The columns correspond to different inclination angles of observation, $i$. The blue line corresponds to the flat disk, the orange line corresponds to the disk with $\dot{M} / \dot{M}_{\text{Edd}} = 0.3$, the green line corresponds to $\dot{M} / \dot{M}_{\text{Edd}} = 0.7$, and the pink line $\dot{M} / \dot{M}_{\text{Edd}} = 1.1$.}
    \label{ellipse_lines}
\end{figure*}

The associated emissivity profiles with these expanded inner disks follow a similar behavior to those of the Shakura-Sunyaev disk. In Figure \ref{ellipse_emis}, we show the emissivity profiles across different coronal heights for the expanded inner disk created from an Eddington ratio of $0.7$. These emissivity profiles show a more extreme shadowing effect on the outer portions of the disk near $100 \, r_\mathrm{g}$ radial along the disk than seen with the Shakura-Sunyaev or flat-disk geometries. The apparent toughs seen in the emissivity profiles near $100 \, r_\mathrm{g}$ are the byproduct of the inner expanded inner disk preventing rays from reaching the outer portion of the expanded inner disk that is decreasing in scale height with increasing radial distance. With increasing coronal height, the amount of self-shielding that occurs reduces. In Figure \ref{ellipse_lines}, we show the line profiles from the expanded inner disk geometry. The behavior of such line profiles is similar to that of the Shakura-Sunyaev disk.

\section{The Warped Disk} \label{warped_sec}
\begin{figure}[htbp] 
    \centering
    \includegraphics[width=\linewidth]{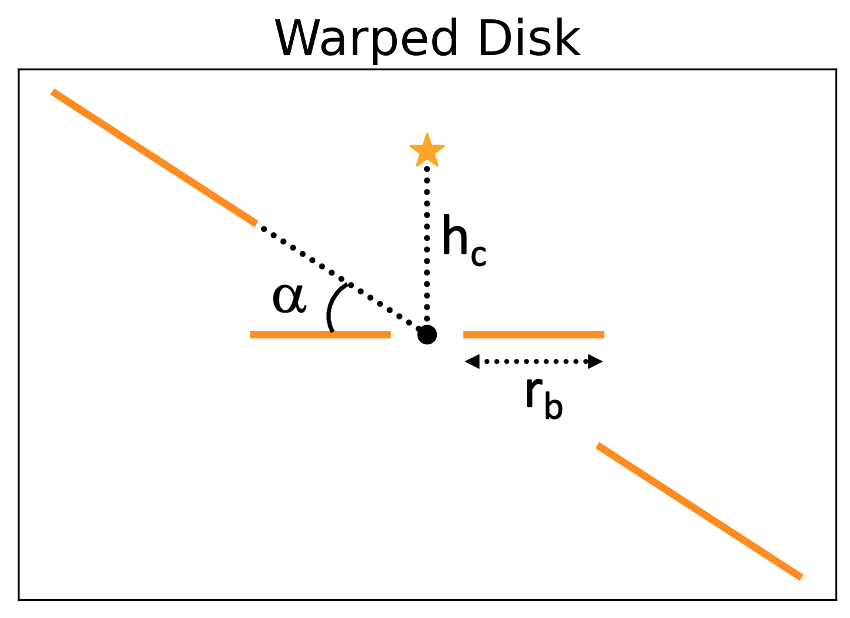}
    \caption{Cross-section of the warped accretion disk geometry viewed edge-on, $\theta = \frac{\pi}{2}$ and $\phi = 0$. Black hole shown at center of the disk geometry and $h_c$ is the height of the lamppost corona above the disk; $r_b$ is the break radius of the inner accretion disk from the outer disk; and $\alpha$ is the angle of misalignment between the inner and outer disk. Note this plot has no associated units and is meant to give an understanding of the basic configuration of the system.}
    \label{warped_disk}
\end{figure}

We investigated a warped disk geometry in which the accretion disk consists of a flat ($\theta = \frac{\pi}{2}$) inner accretion disk that is aligned with the spin axis of the black hole and an outer, misaligned flat disk. We show a schematic of this geometry in Figure \ref{warped_disk}. This type of accretion disk is thought to arise from tidal disruption events involving stellar mass black holes \citet{Tremaine_2014}. This accretion disk geometry was first theorized by \citet{1975ApJ}. The warped disk may be the result of Lense-Thirring precession causing the transition of material in the outer parts of the accretion disk, where the material is misaligned with the spin axis of the black hole, to settle into the equatorial plane of the black hole closer to its innermost stable circular orbit. Additionally, in the context of X-ray binaries, these warped disks could be formed by a supernova kicking a black hole out of alignment with its spin axis \citet{brandt1994effects}. 

We investigate a range of values for the parameters which define the warped disk system including the observation inclination angle, $i$; the break radius of the inner accretion disk, $r_\text{b}$; the angle of misalignment of the outer disk from the inner disk, $\alpha$; the azimuthal angle of the observation, $\phi$; and the height of the corona above the accretion disk, $h_c$. To understand the effects of these various parameters we look at the following values: $r_\text{b} = 10  \, r_\mathrm{g}, \, 20  \, r_\mathrm{g},\, 30  \, r_\mathrm{g} $, $\alpha = 15^\circ,  \,  30^\circ,  \, 45^\circ $, $\phi = 0^\circ - 330^\circ $ in increments of $30^\circ$, and $h_c = 5  \, r_\mathrm{g}, \, 10  \, r_\mathrm{g}, \, 20  \, r_\mathrm{g} $. 

We choose our values for $r_\text{b}$ from \citet{Musoke_2022} where the authors were investigating the wide variability of black hole X-ray binaries, specifically studying the origins of quasi-periodic oscillations (QPOs) observed from these systems. From associated GRMHD simulations, it was found that the inner accretion flow would tear off from the outer accretion disk in warped disk systems at $\sim 10-20 \, r_\mathrm{g}$ for a warped disk with an angle of misalignment of $\alpha = 65^\circ$ and a black hole with spin $a=0.9375$. 

\begin{figure}[htbp] 
    \includegraphics[width=\linewidth]{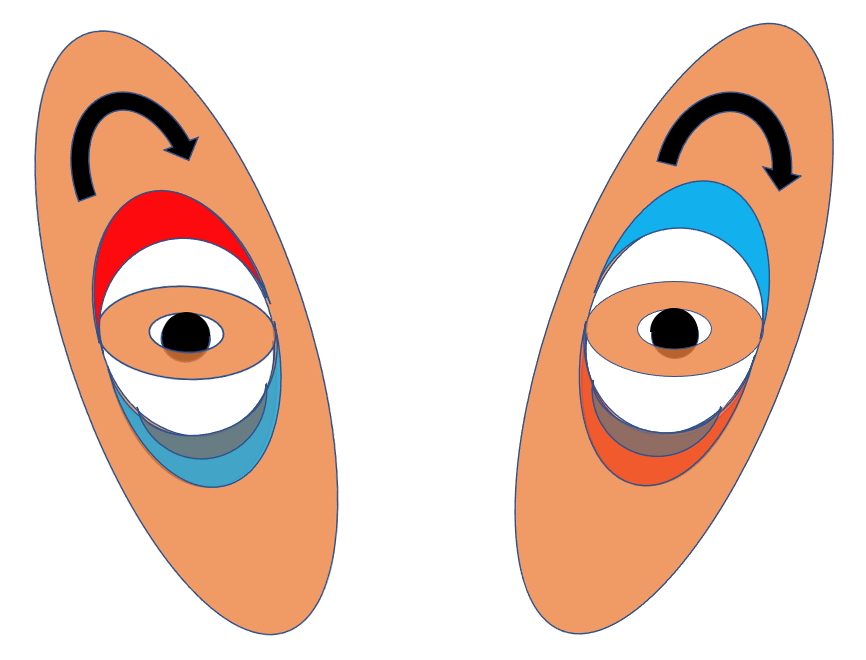}
    \caption{Depiction of the effect of the azimuthal observation angle on a warped disk. From the observer's view, the effective area of the accretion disk is the same. However, because the material in the disk is orbiting in the same direction in both configurations (shown by black arrows), the energy shifts change as a result of the shadowing of the outer disk by the inner disk. On the right, the emitting material moving toward the observer (shown by the shaded in blue region above the black hole) is not shadowed by the inner accretion disk. On the left, this same material is now shadowed by the inner accretion disk, as the inner disk prevents rays from the corona from illuminating the blue region below the black hole. The opposite effect occurs for the emitting material moving away from the observer, shown in red, where the relationship is reversed.}
    \label{azimuthal_depend}
\end{figure}

Our range of $\phi$ values in our simulations covers all the possible azimuthal angles of observation around the disk. This geometry breaks the axial symmetry that is shared by the other models, meaning that the unique azimuthal viewing angles are not encapsulated in $\phi \in \{0^\circ,  \,  30^\circ,  \,  60^\circ,  \, 90^\circ\}$ but rather the entire $360^\circ$ around the spin axis. However, energy shifts across the disk also occur with changes in $\phi$ due to shadowing of the misaligned outer disk by the inner flat accretion flow. For warped-disk configurations that are identical other than being reflected across the z-axis, the reflection spectra will be altered due to these differences in how the material in the disk is moving with respect to the observer. In one of such systems, the material that moves away from the observer will be shadowed by the inner disk, while for the other observation, it will not (see Figure \ref{azimuthal_depend}). 

To create the warped disk geometry in our simulations, we used the following set of stopping conditions in our ray tracing: in the region $r_\text{ISCO} < r < r_b$, we define the inner disk to lie in the equatorial plane of our coordinate system, which is, as usual, aligned with the black hole spin axis and use the condition that $\theta \ge \frac{\pi}{2}$ to stop the propagation of rays. Then, in the region where $0 <\phi\le \pi$ and $r \ge r_b$, the condition was the following: 
\begin{equation}
    y\sin{\alpha}+z\cos{\alpha} \le 0
\end{equation} using Cartesian coordinates with $\alpha$ being the angle of misalignment of the outer disk from the inner flat disk. The break radius $r_b$ is chosen azimuthally such that the outer portion of the disk joins consistently with the inner disk region. For this geometry, we used the same variable step as detailed before.

For the warped disk geometry, the emissivity profiles are a function of $\phi$ and $r$ because there is no longer azimuthal symmetry in this disk geometry. We calculated the associated emissivity profiles by breaking up the warped disk into both azimuthal and radial bins. 

\begin{figure}[htbp] 
    \includegraphics[width=\linewidth]{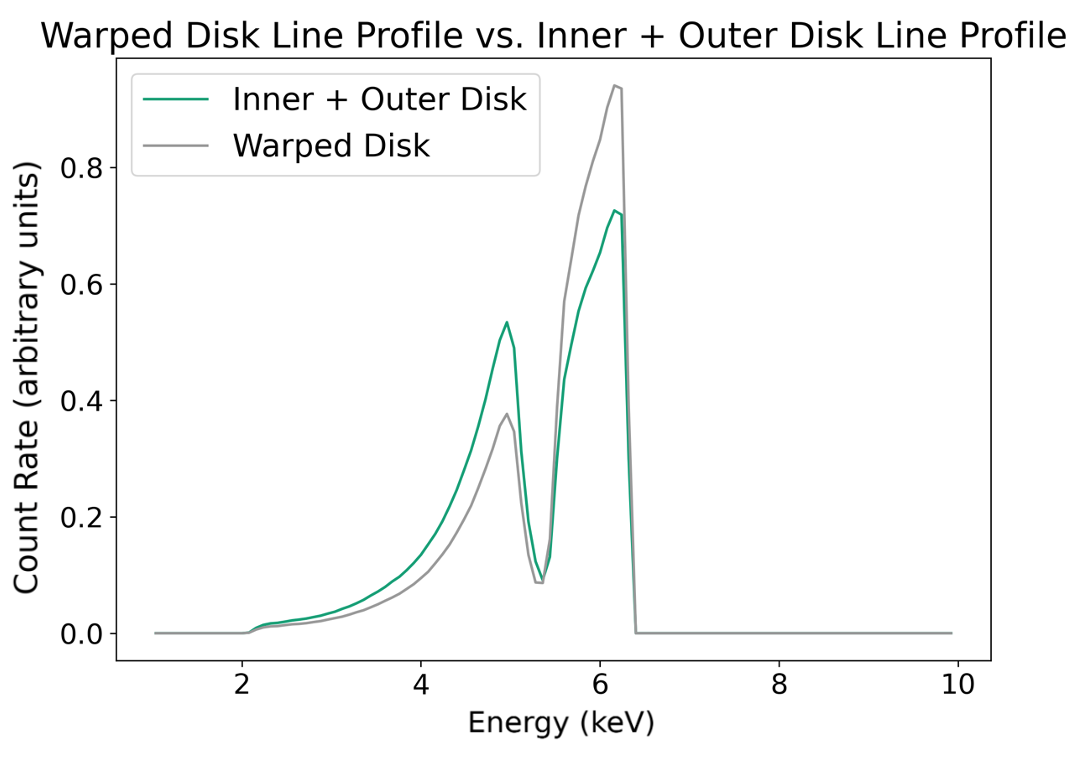}
    \caption{In gray, the line profile from a warped disk with $h_c = 10\, r_\mathrm{g}$, $r_b = 10\, r_\mathrm{g}$, and $\alpha = 30^\circ$ viewed from an inclination of $10^\circ$ and $\phi = 0^\circ$. In green, the line profile from the addition of the line profile from a flat disk with radius $10\, r_\mathrm{g}$ and a misaligned outer disk (the outer portion of the warped disk). Simulating the two disks that make up the warped disk does not give the same line profile as simulating the warped disk in its entirety due to additional energy shifts caused by the warped disk geometry.}
    \label{in_out_warp}
\end{figure}

\begin{figure}[htbp] 
    \centering
    \includegraphics[width=\linewidth]{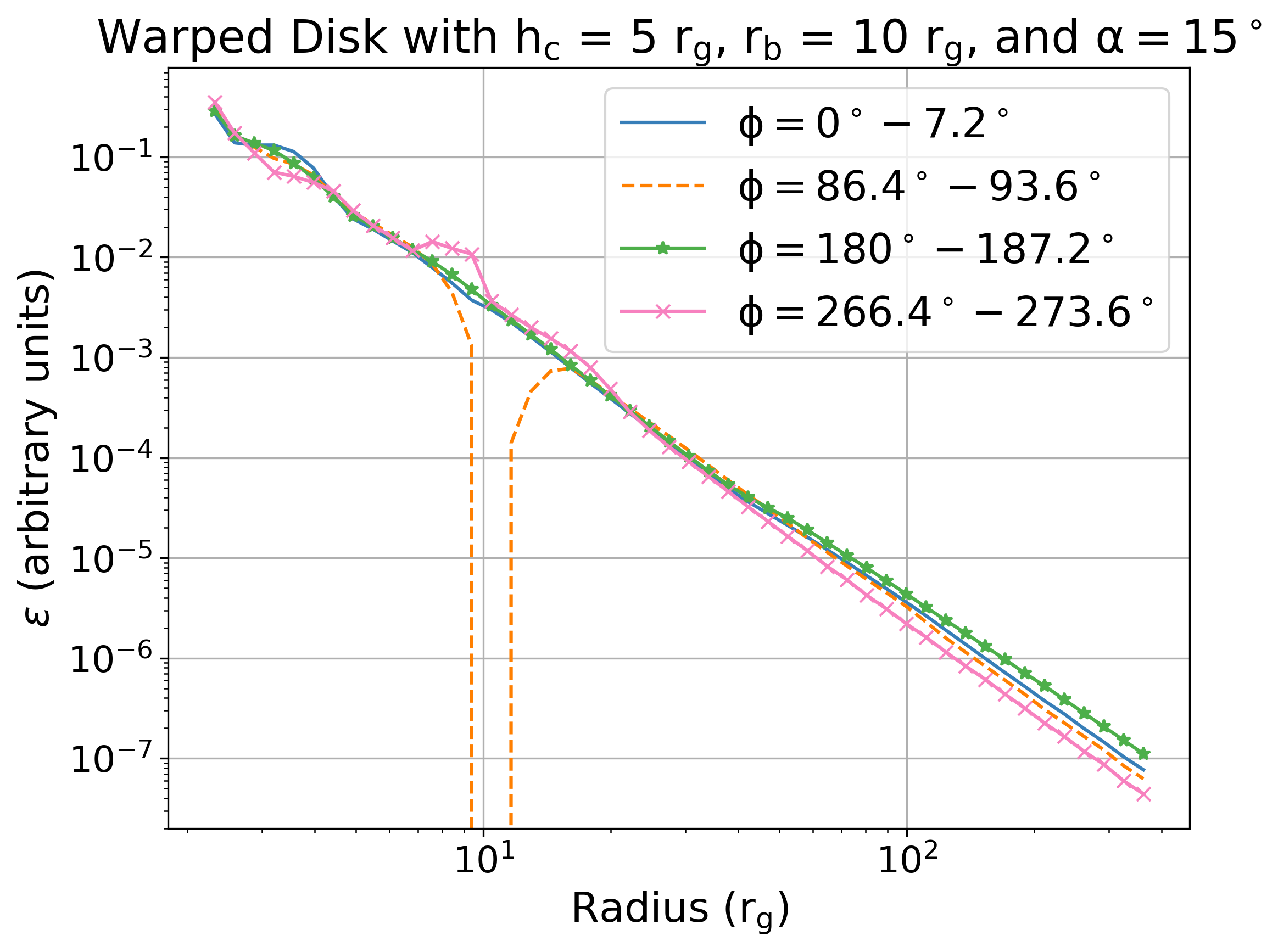}
    \caption{Emissivity profiles associated with a warped accretion disk with $h_c=5\, r_\mathrm{g}$, $r_b=10\, r_\mathrm{g}$, and $\alpha=15^\circ$. Each line corresponds to the same accretion disk geometry but represents the emissivity profile over the displayed range of azimuthal angles $\phi$ around the disk.}
    \label{warped_emi}
\end{figure}

In Figure \ref{warped_disk}, we show the schematic of a warped disk system. In this system, an observation of this geometry along $\phi = 0^\circ$ and $\phi = 180^\circ$ would approximately correspond to the emissivity profile of a flat disk. Additionally, the emissivity profile associated with $0^\circ < \phi < 180^\circ$ would have a reduction in the number of rays that hit the outer disk due to the shadowing effect of the inner disk on the outer disk. This reduction in emissivity reaches a maximum along $\phi = 90^\circ$. In the region with $180^\circ < \phi < 360^\circ$, there is an increase in the number of rays that hit the outer warped disk. This is largely because the corona is physically closer to the outer warped disk. The maximum of this increase in emissivity occurs along with $\phi = 270^\circ$. The position where this reduction or enhancement of emissivity occurs is at $\sim r_\text{break}$. In Figure \ref{in_out_warp}, we see how the warped disk line profiles cannot be created by independently simulating the inner accretion flow (flat disk) and adding the line profile from the outer, misaligned portion of the disk because of the shadowing effect of the inner disk on the outer disk in the full warped disk system. Additionally, in Figure \ref{warped_emi}, we show an example of the resulting emissivity profiles for a warped disk with $h_c=5\, r_\mathrm{g}$, $r_b=10\, r_\mathrm{g}$, and $\alpha=15^\circ$, where these trends are present. 

As with the other disk geometries, with increasing coronal height, the line profiles have more peaked emission. In the context of the warped disk, the height of the corona determines the extent to which the outer accretion disk determines the shape of the line profile. The larger the coronal height, the more the outer warped disk influences the line profile. 

Similarly, increasing the observation inclination angle causes a broadening of the line profile. For the warped disk, however, the effect of inclination is also altered by the azimuthal observation angle. Because of this azimuthal dependence, there are circumstances where, at lower inclination angles, there is more broadening than at higher inclinations. For instance, if the observation has $\phi = 270^\circ$ with an outer disk with a misalignment of $45^\circ$, then observations of the disk become more energy shifted at inclinations around $50^\circ$ than they otherwise would be with any other disk geometry. As a result of this, there are certain observations of systems that become very difficult to make due to a great reduction in the counts of reflected photons, as well as large energy shifts that broaden the line profile.  

\begin{figure}[htbp] 
    \includegraphics[width=\linewidth]{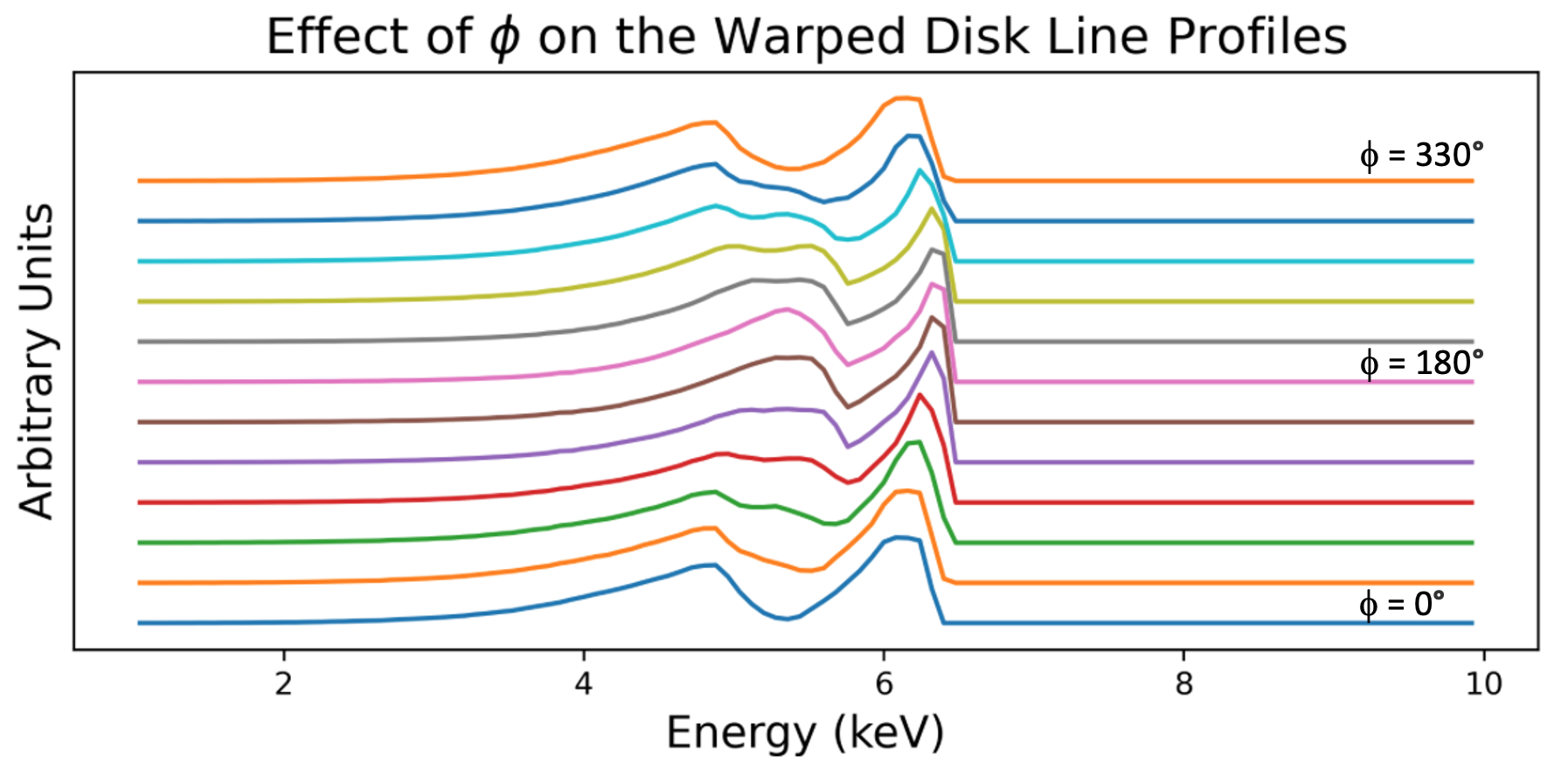}
    \caption{The warped disk line profiles from a warped disk with $r_b = 10\, r_\mathrm{g}$, $h_c = 5 \, r_\mathrm{g}$, and $\alpha = 30^\circ$ viewed from $i=10^\circ$ with varying values of $\phi$. We plot each of the line profiles for each value of $\phi$ from $30^\circ$ to $330^\circ$ on top of each other. We see that the most blueshifted spectra occur around $\phi = 180^\circ$ and the most redshifted line profiles are near $\phi = 0^\circ$. }
    \label{stack_warped_lines}
\end{figure}

Changes in the azimuthal angle of observation shift the location of the peaks in the iron K$\alpha$ line. This relationship is shown in Figure \ref{stack_warped_lines}. The line profiles become most blueshifted when $\phi = 180^\circ$ and most redshifted when $\phi = 0^\circ$ due to the shadowing effects detailed in Figure \ref{azimuthal_depend}.

\begin{figure}[htbp] 
    \includegraphics[width=\linewidth]{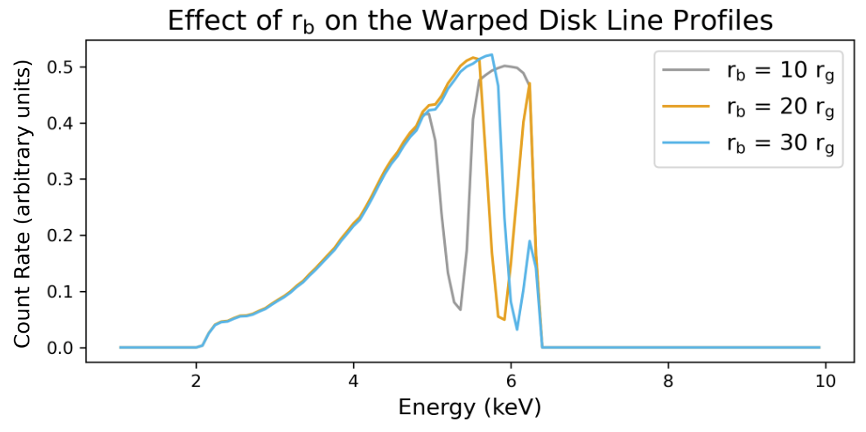}
    \caption{Plot showing the effect of varying the break radius on the warped disk iron K$\alpha$ line. Shown are the line profiles associated with the warped disk with $\alpha = 15^\circ$, $h_c = 5 \, r_\mathrm{g}$ viewed at $i = 10^\circ$ and $\phi=0$. Each line corresponds to a warped disk with a different value of $r_b$. An increase in $r_b$ makes the inner disk more dominant in the line profile. }
    \label{warped_rb}
\end{figure}

Varying the value of the break radius, $r_{\text{b}}$, influences how much the inner flat accretion disk determines the shape of the line profile. With larger values for $r_{\text{break}}$, the more the iron K$\alpha$ line is dominated by the inner flat disk. This is best seen at low inclination angles as in Figure \ref{warped_rb}. This behavior causes the line profiles in certain observations to exhibit less of a doubly peaked structure (usually in the case of small angles of misalignment).  

\begin{figure}[htbp] 
    \centering
    \includegraphics[width=\linewidth]{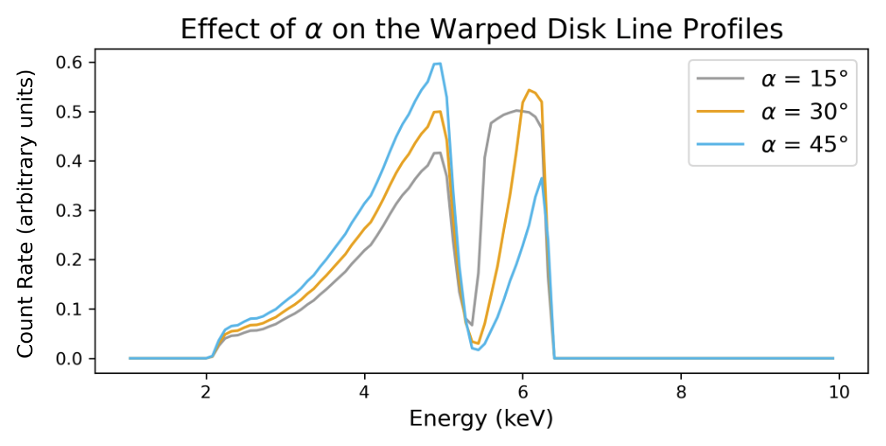}
    \caption{Line profiles from warped disks with $r_b = 10\, r_\mathrm{g}$, $h_c = 5\, r_\mathrm{g}$, viewed from $i = 10^\circ$ and $\phi=0^\circ$ and varying angles of misalignment, $\alpha$. Increasing $\alpha$ increases the shadowing effect from the inner disk on the misaligned disk, as seen in the reduction in the counts of photons between the doubly peaked reflection spectra. }
    \label{warped_mis}
\end{figure}

Additionally, the angle of misalignment, $\alpha$, of the outer disk and the inner flat disk affects the shape of the line profiles. Larger misalignment angles at low observation inclinations correspond to more energy shifting in reflection spectra. This is because higher angles of misalignment at low inclinations correspond to the outer disk being viewed as a flat disk at more of an inclined observation. Thus, there is an increase in the number of the most redshifted photons as well as a decrease in the counts of moderately redshifted and blueshifted rays seen in the line profiles in Figure \ref{warped_mis}. 

In addition to this, the effects detailed in Figure \ref{azimuthal_depend} can be further enhanced by an increase in this warp angle. Namely, a greater angle of misalignment causes more shadowing of the outer disk by the inner disk. This can be seen in Figure \ref{warped_mis} by the continued reduction in the counts of photons between the doubly peaked line profiles as $\alpha$ is increased. 

\begin{figure*}
    \centering
    \includegraphics[width=\linewidth]{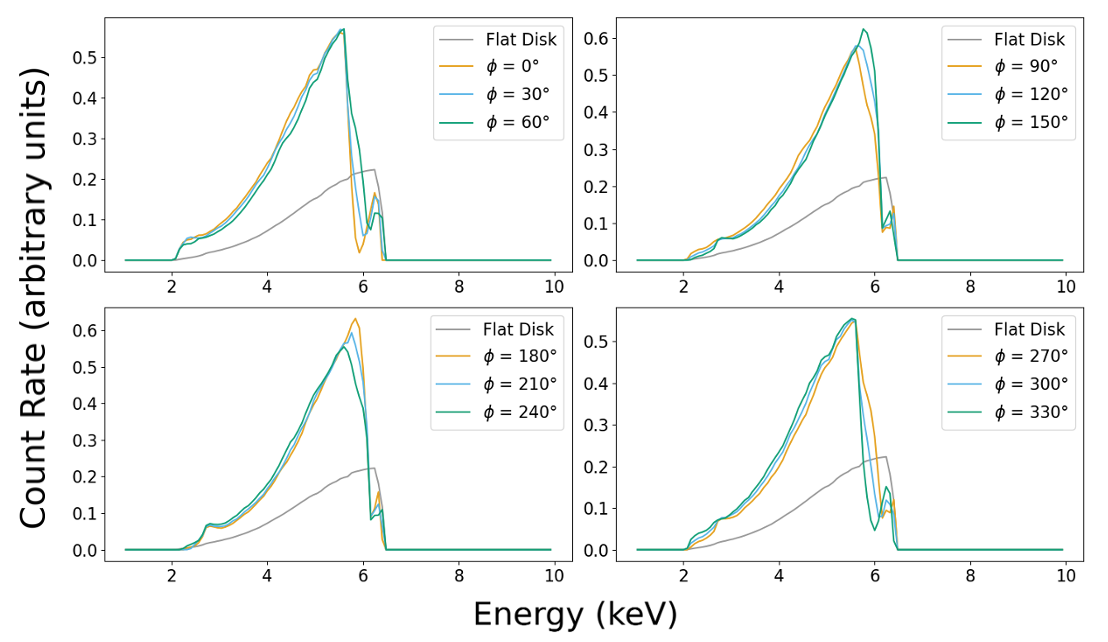}
    \caption{The flat-disk line profile (gray line) plotted against all the line profiles from the warped disk with $r_b = 20\, r_\mathrm{g}$, $\alpha = 30^\circ$, and $h_c = 5\,r_\mathrm{g}$ across all values of $\phi$. Each panel corresponds to a different set of $\phi$ values.}
    \label{warped_v_flat}
\end{figure*}

Now, the line profiles associated with the warped disk consistently differ in their overall behavior from those of the flat disk. In Figure \ref{warped_v_flat}, we show a collection of the warped disk line profiles versus those from the flat disk. The warped disk geometries discussed here can all be distinguished with XRISM from the flat disk, as well as the other disk geometries, given the unique behavior of their line profiles. The magnitude of the effect in the line profiles from the warped disk geometry is sensitive to the break radius of the warped disk. From Figure \ref{warped_rb}, we can see that for a break radius of $10\, r_\mathrm{g}$ the effect is very detectable, but as the break radius is increased to $20\, r_\mathrm{g}$ and $30\, r_\mathrm{g}$, the effect of the warp becomes less prominent. For smaller break radii, the effects from the warped disk geometry are so prevalent that they may be seen even without the resolution of XRISM.  

\begin{deluxetable*}{lcccccl}
\tablecaption{Mean signed differences from each valid fit of the flat-disk model to the constant-aspect-ratio disk with different aspect ratios (rows) and different coronal heights (columns). In each entry, there is the mean signed difference calculated for the spin, coronal height, and inclination angle (in order from top to bottom) with the associated error. We additionally report the smallest reduced $\chi^2$ values for each system and report those values in the rows labeled, ``Minimum $\chi^2_\text{red}$." Here, * indicates that the mean signed difference was not statistically significant from a t-test for bias with regard to the true parameter value in the constant-aspect-ratio disk system. Note that for the case of the constant-aspect-ratio disk with $\frac{h}{\rho}=1$ and $\text{h}_\text{c}=2.5\, \text{r}_\text{g}$ the flat-disk model did not fit the constant-aspect-ratio disk system, indicated by the use of `N/A.'}
\tablehead{
\colhead{} & \colhead{Parameter} & \colhead{$\text{Units}$} & \colhead{$h_c=2.5\, r_\mathrm{g}$} & \colhead{$h_c=5\, r_\mathrm{g}$} & \colhead{$h_c=10\, r_\mathrm{g}$} & \colhead{$h_c=20\, r_\mathrm{g}$}}
\startdata
\multirow{3}{*}{$\frac{h}{\rho}=0.1$} & Spin & $GMc^{-2}$ & $-0.01 \pm 0.04^*$ & $-0.01 \pm 0.06^*$ & $-0.14 \pm 0.18$ & $-0.23 \pm 0.26$ \\
& coronal height& $\text{r}_\text{g}$ & $-0.35\pm 0.43$ & $-0.43 \pm 0.41$ & $-1.1 \pm 1.0$ & $-0.9 \pm 0.9$ \\
& Inclination& $\text{deg}$ & $-1.4 \pm 1.3$ & $-1.2 \pm 1.0$ & $-0.6 \pm 0.8$ & $-0.8 \pm 0.8$ \\
& Minimum $\chi^2_\text{red}$ & & $1.031$ & $0.977$ & $0.968$ & $0.974$ \\
\hline
\multirow{3}{*}{$\frac{h}{\rho}=0.5$} & Spin & $GMc^{-2}$ & $0.01\pm 0.03^*$ & $-0.26 \pm 0.25$ & $-0.18 \pm 0.24$ & $-0.35 \pm 0.23$ \\
& coronal height&$\text{r}_\text{g}$ & $-0.9\pm 0.3$ & $-2.0 \pm 1.4$ & $-0.9 \pm 0.7$ & $-2.0 \pm 1.1$ \\
& Inclination& $\text{deg}$ & $-5 \pm 3$ & $-4 \pm 4$ & $-4.3 \pm 1.8$ & $-3.4 \pm 1.1$ \\
& Minimum $\chi^2_\text{red}$  &  & $1.071$ & $0.992$ & $1.076$ & $1.015$ \\
\hline
\multirow{3}{*}{$\frac{h}{\rho}=1$}  & Spin & $GMc^{-2}$ & N/A & $-0.36 \pm 0.13$ & $-0.42 \pm  0.20$ & $-0.33 \pm 0.28$ \\ 
& coronal height& $\text{r}_\text{g}$ & N/A & $-3.2 \pm 0.4$ & $-1.4 \pm 0.3$ & $-1.2 \pm 0.9$ \\
& Inclination& $\text{deg}$ & N/A & $-10.1 \pm 3.0$ & $-10.7 \pm 2.1$ & $-7.5 \pm 2.8$ \\
& Minimum $\chi^2_\text{red}$ & & N/A & $1.205$ & $1.413$ & $1.279$ \\
\enddata
\end{deluxetable*}\label{tab:wedged_parameters}

\section{Discussion} \label{discussion}
We have investigated the effects of accretion disk geometry on the iron K$\alpha$ line with new general relativistic ray tracing simulations. We survey a range of different disk geometries, including the constant-aspect-ratio disk, the Shakura-Sunyaev disk, the expanded inner disk, and the warped disk. We find that with the resolution from a real XRISM observation, using a flat-disk model to fit line profiles created for systems with accretion disk geometry could have noticeable changes in estimates of the coronal height, spin, and inclination angle of the system. In less extreme accretion disk scenarios such as a constant-aspect-ratio disk with $\frac{h}{\rho} = 0.1$ or a Shakura-Sunyaev disk or inner expanded disk with $\dot{M} / \dot{M}_{\text{Edd}} = 0.3$, the flat-disk model provides a reasonable fit to the line profiles associated with these accretion geometries. However, as the accretion disk geometry becomes more extreme with $\frac{h}{\rho} = 0.5$ and $1$ for the constant-aspect-ratio disk and $\dot{M} / \dot{M}_{\text{Edd}} = 0.7$ and $1.1$ for the Shakura-Sunyaev disk or inner expanded disk, using a flat-disk model results in underestimates in coronal height and inclination angle. Estimates of black hole spin may be affected as well for the case of $\frac{h}{\rho} = 1$ or $\dot{M} / \dot{M}_{\text{Edd}} = 1.1$, though the relationship is not as clear for the intermediate disk thicknesses of $\frac{h}{\rho} = 0.5$ or $\dot{M} / \dot{M}_{\text{Edd}} = 0.7$. Additionally, the warped disks that we simulated could not be adequately fit with the flat-disk model. This means that we have the ability to better estimate coronal height, spin, and inclination angle in extreme accretion scenarios, as well as detect the warped disk accretion geometry observationally. 

\subsection{Effect of Geometry on Parameter Estimation} \label{geometry_effects}
In order to determine if this variation in geometry can be detected in the spectrum, we compare the line profiles from the disk with a non-trivial geometry (i.e., constant-aspect-ratio, Shakura-Sunyaev, expanded inner disk) to those from a flat disk. We use the $\chi^2$ statistic to determine whether the flat-disk model provides an adequate description of the simulated non-trivial disk geometry spectrum. We consider the iron line spectra measured with the microcalorimeter resolution and signal-to-noise similar to that of XRISM Resolve. Each spectrum is binned with $30$ eV bins with an associated error of $5\%$ of the counts within each energy bin. We set the uncertainty in each spectral bin to correspond to the count rate detected from the nearby AGN MCG-6-30-15 \citet{brenneman2025}. The spectral resolution of the XRISM Resolve microcalorimeter spectrometer is $5$ eV; however, we bin to $30$ eV to enhance the signal-to-noise when observing a system with a flux level typical of nearby AGN such as MCG-6-30-15. The model spectrum consists of the line and the primary continuum emission from the corona, modeled by a power law with a photon index of $2$. This step is taken to ensure that the small differences in the counts in energy bins with few rays do not substantially affect our $\chi^2$ calculation.

First, the spectra from the constant-aspect-ratio disk with $a = 0.9\,  GMc^{-2}$, $\frac{h}{\rho} = 0.1, \, 0.5, \, 1$, $\text{h}_\text{c} = 2.5 \, \text{r}_\text{g}, \, 5 \, \text{r}_\text{g}, \, 10 \, \text{r}_\text{g}, \, 20 \, \text{r}_\text{g}$, and $i = 10^\circ, \, 20^\circ, \, 30^\circ, \, 40^\circ,\, 50^\circ, \, 60^\circ, \, 70^\circ, \, 80^\circ,$ are compared to a grid of flat-disk line profiles which we interpolate over with parameter ranges of $a = 0.2 - 0.99 \, GMc^{-2}$, $\text{h}_\text{c} = 1 \, \text{r}_\text{g} - 25 \, \text{r}_\text{g}$, and $i = 4^\circ - 80^\circ$. We find the flat-disk model parameters that best match the simulated data by minimizing $\chi^2$. A model is considered an acceptable fit if we cannot reject the null hypothesis at the $5 \%$ significance level (a $p$-value greater than $0.05$). To quantify the bias in the fits, we calculate the mean signed difference of the values that are acceptable under our criterion and then perform a t-test to assess if there is statistically significant bias. The resulting mean signed differences of this process are summarized in Table \ref{tab:wedged_parameters}. We additionally report the minimum reduced $\chi^2$ values obtained for each set of disk parameters in Table \ref{tab:wedged_parameters}. 

For the constant-aspect-ratio disk with $\frac{h}{\rho} = 0.1$ with coronal height $2.5 \, r_\mathrm{g}$, we find no significant deviations in the estimates of black hole spin and coronal height between the constant-aspect-ratio disk and flat-disk models. In the first two entries of the first column of Table \ref{tab:wedged_parameters}, we see that the accepted value (parameter value of constant-aspect-ratio disk model) falls within the uncertainty, making the values consistent of spin and coronal height consistent between the models. This is seen by simply comparing the magnitude of the mean sign difference value to the uncertainty. If the uncertainty is larger than the magnitude of the mean signed difference, then the accepted value falls within the uncertainty. The inclination estimate is very slightly underestimated as we find that the value of the average mean signed difference is $-1.4^\circ \pm 1.3^\circ$. For the larger coronal heights of $5 \, r_\mathrm{g}$, $10 \, r_\mathrm{g}$, and $20 \, r_\mathrm{g}$, we find no significant deviations in the spin but underestimates of coronal height as well as inclination angle. These underestimates in coronal height values are smallest when the coronal height is $5 \, r_\mathrm{g}$, being $\sim 0.4 \, r_\mathrm{g}$. Additionally, for coronal heights of $10$ and $20 \, r_\mathrm{g}$ the underestimates are roughly in the range of $\sim 1\, r_\mathrm{g}$. The underestimates in the inclination are small, being roughly a degree or so different (note that for $h_c=10 \, r_\mathrm{g}$ there is no significant underestimate of the inclination). 

As the aspect ratio of the constant-aspect-ratio disk increases, these underestimates become more prominent. For the constant-aspect-ratio disk with $\frac{h}{\rho} = 0.5$, we find there is no significant deviations in the estimates from between the constant-aspect-ratio disk model and flat-disk model for coronal heights of $2.5 \, r_\mathrm{g}$ and $10 \, r_\mathrm{g}$ as the accepted spin values fall within the uncertainty. Additionally, for $h_c=5 \, r_\mathrm{g}$, the accepted value falls very close ($\sim 0.01 \, GMc^{-2}$) to being within the uncertainty, making this a slight bias in the parameter estimation using the flat-disk model. For the $h_c=20 \, r_\mathrm{g}$, the underestimation of the coronal height is more definitive with the mean signed difference being $-0.35 \pm 0.23\, GMc^{-2}$. Additionally, the coronal heights and inclination angles of the flat-disk model are more underestimated when fit to this constant-aspect-ratio disk with $\frac{h}{\rho} = 0.5$. In order of increasing coronal height above the constant-aspect-ratio disk, the mean signed differences of the flat-disk fits from the constant-aspect-ratio disks parameters are $-0.9 \pm 0.3\, r_\mathrm{g}$, $-2.0 \pm 1.4\, r_\mathrm{g}$, $-0.9 \pm 0.7\, r_\mathrm{g}$, and $-2.0 \pm 1.1\, r_\mathrm{g}$ for the coronal height and $-5^\circ \pm 3^\circ$, $-4^\circ \pm 4^\circ$, $-4.3^\circ \pm 1.8^\circ$, and $-3.4^\circ \pm 1.1^\circ$ for the inclination angles. 

These trends continue for the constant-aspect-ratio disk with the largest aspect ratio that we analyze. For the constant-aspect-ratio disk with $\frac{h}{\rho} = 1$, the flat-disk model does not provide an acceptable fit for the constant-aspect-ratio disk with $h_c = 2.5 \, r_\mathrm{g}$. For the rest of the coronal heights, the flat disk does fit the constant-aspect-ratio disk line profiles. We find large underestimates of the spin, coronal height, and inclination angles of the flat disks that fit the constant-aspect-ratio disk line profiles. The associated mean signed difference on the spin of the flat-disk models that fit the constant-aspect-ratio disk line profiles are $-0.36 \pm 0.13 \, GMc^{-2}$, $-0.42 \pm 0.20 \, GMc^{-2}$, and $-0.33 \pm 0.28 \, GMc^{-2}$ in order of increasing coronal height above the constant-aspect-ratio disk. For the mean signed differences for coronal height, these respective values are $-3.2\pm 0.4\, r_\mathrm{g}$, $-1.4 \pm 0.3\, r_\mathrm{g}$, and $-1.2 \pm 0.9\, r_\mathrm{g}$ and $-10.1^\circ \pm 3.0^\circ$, $-10.7^\circ \pm 2.1^\circ$, and $-7.5^\circ \pm 2.8^\circ$ for the inclination angles. 

Generally, the estimates of the spin and coronal height of the flat disks decrease as the height of the corona above the constant-aspect-ratio disk increases. This effect occurs for two reasons. First, a flat-disk model with lower spin and lower coronal height can have similar behavior as a flat disk with higher spin and a larger coronal height \citet{2012MNRAS.424..217F, 2014MNRAS.439.2307F}. Additionally, the thickness of the constant-aspect-ratio disk causes an enhancement in the line profiles at intermediate redshift values of $\sim 4-6$ keV and a slight decrease in more redshifted photons. This causes further underestimates of spin and coronal height for the flat-disk models to fit these features.  

We performed this same analysis but fit the other disk geometries to a set of constant-aspect-ratio disks. We simulate a new set of constant-aspect-ratio disks where we define the aspect ratio of the constant-aspect-ratio disk to be such that the constant-aspect-ratio disk intersects the non-constant-aspect-ratio disk (i.e., Shakura-Sunayev or the expanded inner disk) at the point of half of the cumulative reflected flux of the non-constant-aspect-ratio disk. We hold the spin of these constant-aspect-ratio disks constant and explore fitting these disks to the line profiles of the non-constant-aspect-ratio disk. We find that the constant-aspect-ratio disk fits the Shakura-Sunayev disk as well as the expanded inner disk for the lower Eddington ratios of $\dot{M} / \dot{M}_{\text{Edd}} = 0.3$ and $0.7$. However, for the thickest of the disks with $\dot{M} / \dot{M}_{\text{Edd}} = 1.1$, there are more notable biases in the parameter estimations when one fits the constant-aspect-ratio disk to the Shakura-Sunyaev disk or expanded inner disk model.

In summary, the flat-disk model provides an adequate fit to line profiles with less extreme accretion disk geometries. Modeling the effects of accretion disk geometry becomes more important when disk thickness becomes more extreme, with aspect ratios of $\frac{h}{\rho} = 0.5$ and $1$. In the case of an accretion disk with $\frac{h}{\rho} = 0.5$, not accounting for this disk geometry may still leave estimates of black hole spin unaffected; however, the estimates of coronal height and inclination angle will be changed. For the most extreme geometry with $\frac{h}{\rho} = 1$, all the parameters are noticeably underestimated if the flat-disk model is used to fit the disk. Additionally, this trend holds for the Shakura-Sunayev and the expanded inner disk in that the modeling disk geometry becomes important at the more extreme accretion scenarios of $\dot{M} / \dot{M}_{\text{Edd}} = 0.7$ and $1.1$. 

\subsection{Additional Remarks} In the ray tracing simulations used in this paper, we simulate the profile of a single relativistically broadened emission line. This is most clearly seen in the iron K$\alpha$ line in the spectrum, and is the kernel with which the rest frame reflection spectrum is convolved. With thick accretion disk geometries with high Eddington ratios, the ionization of the accretion flow, as well as the gradients in the ionization with respect to radius, would additionally be important in determining the shape of the full reflection spectrum. We do not perform these calculations in this paper and defer them for future work. 

Additionally, in this analysis, we did not include the effect of the outward velocity of the material in these accretion disk geometries. We assumed that the material in each of the disks follows Keplerian orbits. By including the effect of the outward velocity of material from the disk, there may be additional alterations to the line profiles discussed previously. In particular, this effect would become important for disk geometries with super-Eddington accretion rates. Recent GRMHD simulations have consistently demonstrated that super-Eddington disks generate optically and geometrically thick winds due to the significant radiation pressure within the disk \citet{Ohsuga_2009}; \citet{Jiang_2014}; \citet{McKinney_2014}; \citet{2014MNRAS.439..503S}. The effects of these winds on the line profiles of these systems have been investigated by \citet{Thomsen_2019}. The most prominent difference seen with the inclusion of the outward velocity of the winds toward the observer is a large blueshifting of the iron K$\alpha$ line. Additionally, in our simulations, we did not include returning radiation, i.e., rays that reflect off of the disk and return to it, getting reflected multiple times before being seen by the observer \citet{1976ApJ...208..534C}. The future inclusion of this effect could be significant for all accretion geometries, as it was previously found to significantly affect the flat-disk model \citet{Wilkins_2020}.

This paper contributes to the continually growing investigation of the iron K$\alpha$ line and all the information that we can derive about black holes from it. Using the simulations detailed above, we can start to understand the geometry of the accretion disks around black holes. In future observations with XRISM, we will collect data on even more luminous sources of X-ray reflection, such as X-ray binaries. These observations will have greater spectral resolution, which will in turn help us further distinguish the finer features of the X-ray reflection spectrum and the iron K$\alpha$ line that are altered by the geometry of the accretion disks associated with these systems. With even better X-ray telescopes (i.e., Athena) on the horizon, the ability to observe accretion disk geometry via the reflection spectrum will only continue to improve. 

\section{Conclusion} \label{conclusion}
We have produced and analyzed a new suite of general-relativistic ray-tracing simulations investigating the effect of accretion disk geometry on the iron K$\alpha$ line. We find that the constant-aspect-ratio disk, the Shakura-Sunyaev disk, and the expanded inner disk caused underestimates in black hole spin, coronal height, and inclination angle when the flat-disk model was fit to the line profiles associated with each of these geometries. The effects of each geometry were most extreme with the geometries with the aspect ratios of $\frac{h}{\rho}=0.5$ and $1$ (in the case of the constant-aspect-ratio disk) or the highest Eddington ratios ($\dot{M} / \dot{M}_{\text{Edd}} = 0.7$ and $1.1$ for the Shakura-Sunyaev disk and expanded inner disk). These underestimates can be very significant with the values in the spin estimates being roughly in the range of $0.5-0.7 \, GMc^{-2}$ when the accepted value is $0.9 \, GMc^{-2}$. We find that only at the most extreme coronal height and Eddington ratios can one potentially distinguish the Shakura-Sunyaev and the expanded inner disks from the constant-aspect-ratio disk geometry. The warped disk caused significant alterations to the iron K$\alpha$ line. The effect of this geometry is sensitive to the size of the break radius, $r_b$. Greater angles of misalignment and smaller break radii enhance the effect of the warped-disk geometry. Additionally, the azimuthal angle of observation directly influences the energy shift of the line profile. In all cases, accretion disk geometry had a notable effect on the iron K$\alpha$ line. With XRISM observations, we can start to develop a better understanding of accretion disk geometry with the inclusion of these disk models. 

\begin{acknowledgments} 
The authors thank Steven Allen at Stanford University for his support and guidance. W.S. acknowledges support from the Stanford Major Grant Program, which provided funding for this project while at Stanford University. 
\end{acknowledgments}

%

\vspace{5mm}


\software{MATPLOTLIB \citet{Hunter:2007}, SCIPY \citet{2020SciPy-NMeth}}





\bibliography{references}{}
\bibliographystyle{aasjournal}



\end{document}